\documentclass{svmult}
\usepackage{makeidx}         
\usepackage{graphicx}        
\usepackage{multicol}        
\usepackage[bottom]{footmisc}
\makeindex             
\begin{document}

\title*{The Macro Model of the Inequality Process and The
Surging Relative Frequency of Large Wage Incomes}
\titlerunning{The Macro Model of the Inequality Process}
\author{John Angle}
\institute{Inequality Process Institute,\\
Post Office Box 429, Cabin John, Maryland, 20818, USA.\\
\texttt{angle@inequalityprocess.org}}

\maketitle
\begin{small}
Revision and extension of a paper, 
`U.S. wage income since 1961: the perceived inequality trend', 
presented to the annual meetings of the Population Association of America, 
March-April 2005, Philadelphia, Pennsylvania, USA. On-line at:
http://paa2005.princeton.edu/download.aspx?submissionID=50379.
\end{small}

\section{The Surge In Wage Income Nouveaux Riches in the U.S., 1961-2003}
Angle (2006) shows that the macro model of the
Inequality Process provides a parsimonious fit to the U.S.
wage income distribution conditioned on education, 1961-2001.
Such a model should also account for all time-series of scalar
statistics of annual wage income. The present paper examines one
such time-series, the relative frequency of large wage incomes
1961-2003. Figure 1 shows an aspect of this kind of statistic: the
larger the wage income, the greater the proportional increase in
its relative frequency. The phenomenon that figure 1 shows, a
surge in wage income \underline{nouveaux riches}, has caused some alarm
and given rise to fanciful theories. The present paper shows that
the macro model of the Inequality Process (IP) accounts for this
phenomenon. In fact, it is simply an aspect of the way wage
income distributions change when their mean and all their
percentiles increase, which they do simultaneously, i.e., it is good
news for a much larger population than the \underline{nouveaux riches}
alone.
\begin{figure}
\centering
    {\resizebox{6.0cm}{!}{\includegraphics{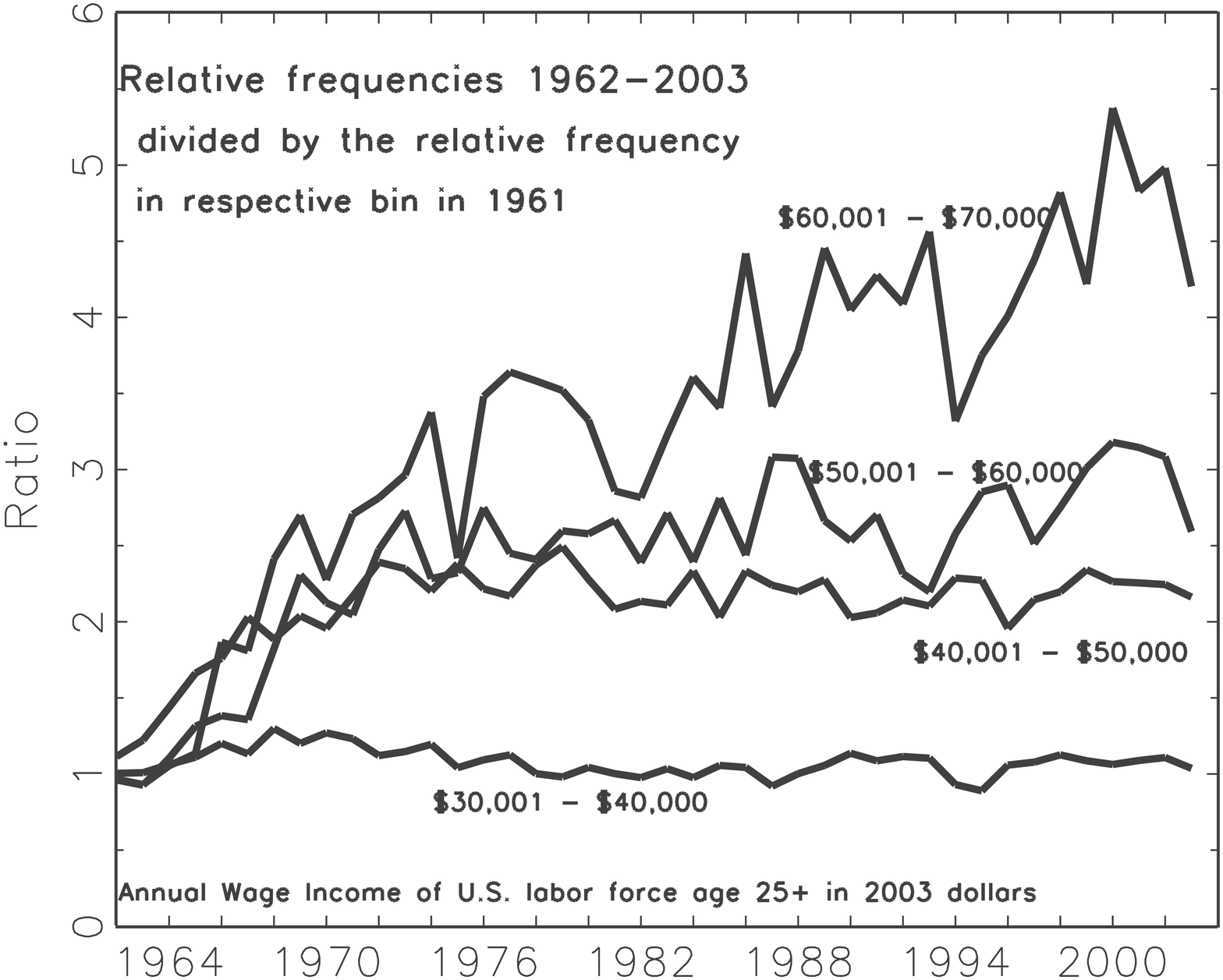}}}
    {\resizebox{6.0cm}{!}{\includegraphics{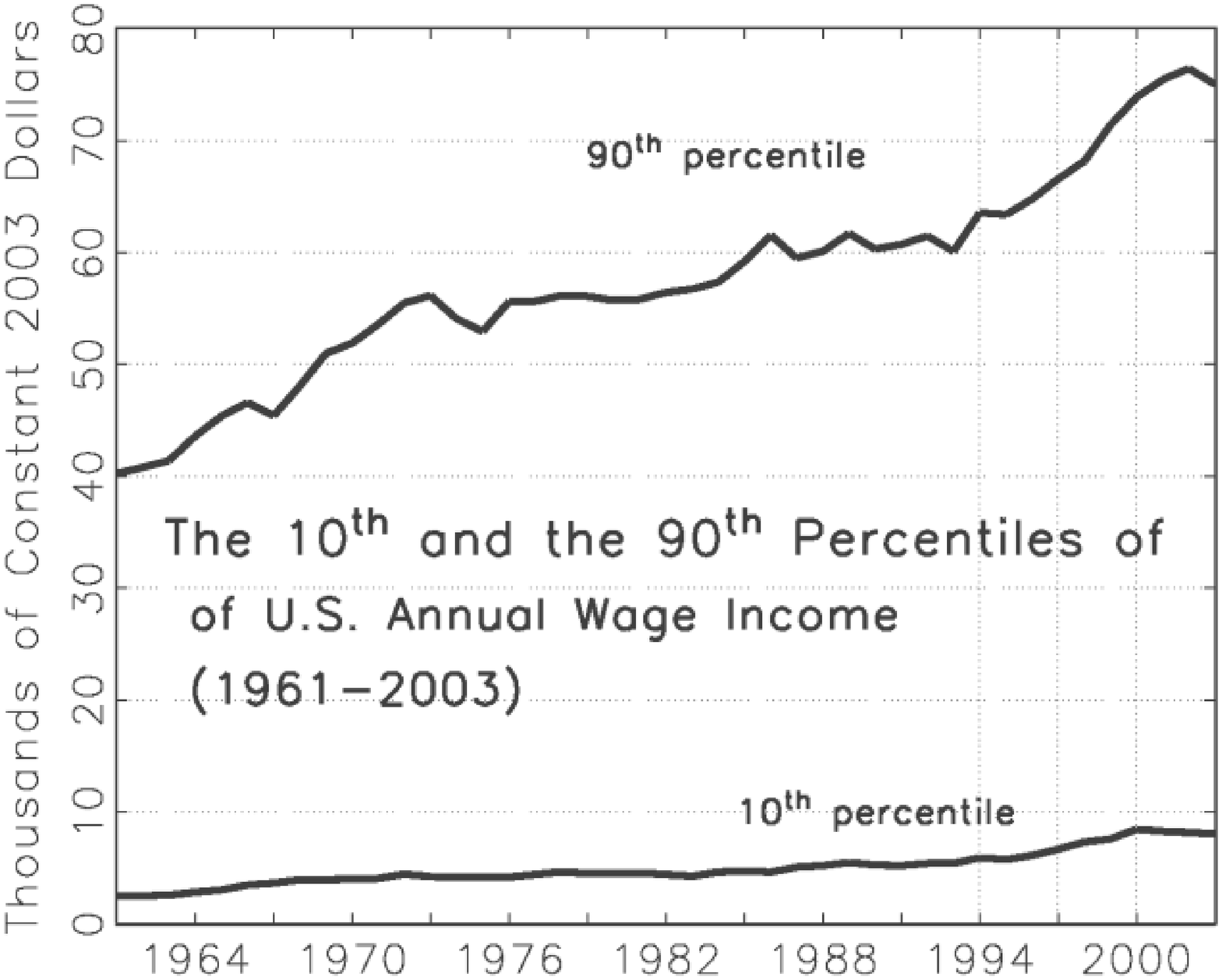}}}
\\    Fig. 1 \hskip 2in Fig. 2
\caption{\label{jafig:fig1}
Ratio of relative frequencies $1962-2003$ in wage income
bins \$1-\$10,000, \$10,001-\$20,000 etc. in terms of constant 2003
dollars to the relative frequency in each bin in 1961
Source: Author's estimates of data of the March Current
Population Survey.
}
\caption{\label{jafig:fig2}
Source: Author's estimates from March CPS data.
}
\end{figure}
Nevertheless, many economists and sociologists have
interpreted the surge in wage income \underline{nouveaux 
riches}\footnote{The term \underline{nouveaux riches} perhaps brings to 
mind the new wealth of entrepreneurs most of whose income is from
tangible assets. \underline{Nouveaux riches} is used here only 
to name the earners of wage income who have begun to earn a wage income
much larger than the average.}
as an
alarming bifurcation of the U.S. wage income distribution into
two distributions, one poor, the other rich, a `hollowed out' 
distribution. Fear of the `hollowing out' of the U.S. wage income
distribution has not only roiled academia but has resulted in alarmed 
press reports and even become an issue in the 2004 U.S.
presidential campaign.

The present paper shows that an increase in mean wage income decreases 
the relative frequency of wage incomes smaller
than the mean and increases the relative frequency of 
wage incomes greater than the mean. Distance from the mean of a
particular wage income, call it $x_0$, is a factor in how fast 
the relative frequency of wage incomes of that size change. For $x_0$'s
greater than the mean, the greater $x_0$, the greater the 
proportional growth in its relative frequency. There is an analogous and
compensating decrease in the relative frequency of wage incomes 
smaller than the mean. The IP's macro model implies that
the wage income distribution stretches to the right when the unconditional
mean of wage income increases, explaining both the surge in wage
income \underline{nouveaux riches} and the fact that the bigger wage income 
percentile has grown more than the smaller wage income
percentile. Data on U.S. wage incomes 1961-2003 confirm the 
implications of the IP's macro model.
\begin{figure}
\centering
    {\resizebox{6.0cm}{!}{\includegraphics{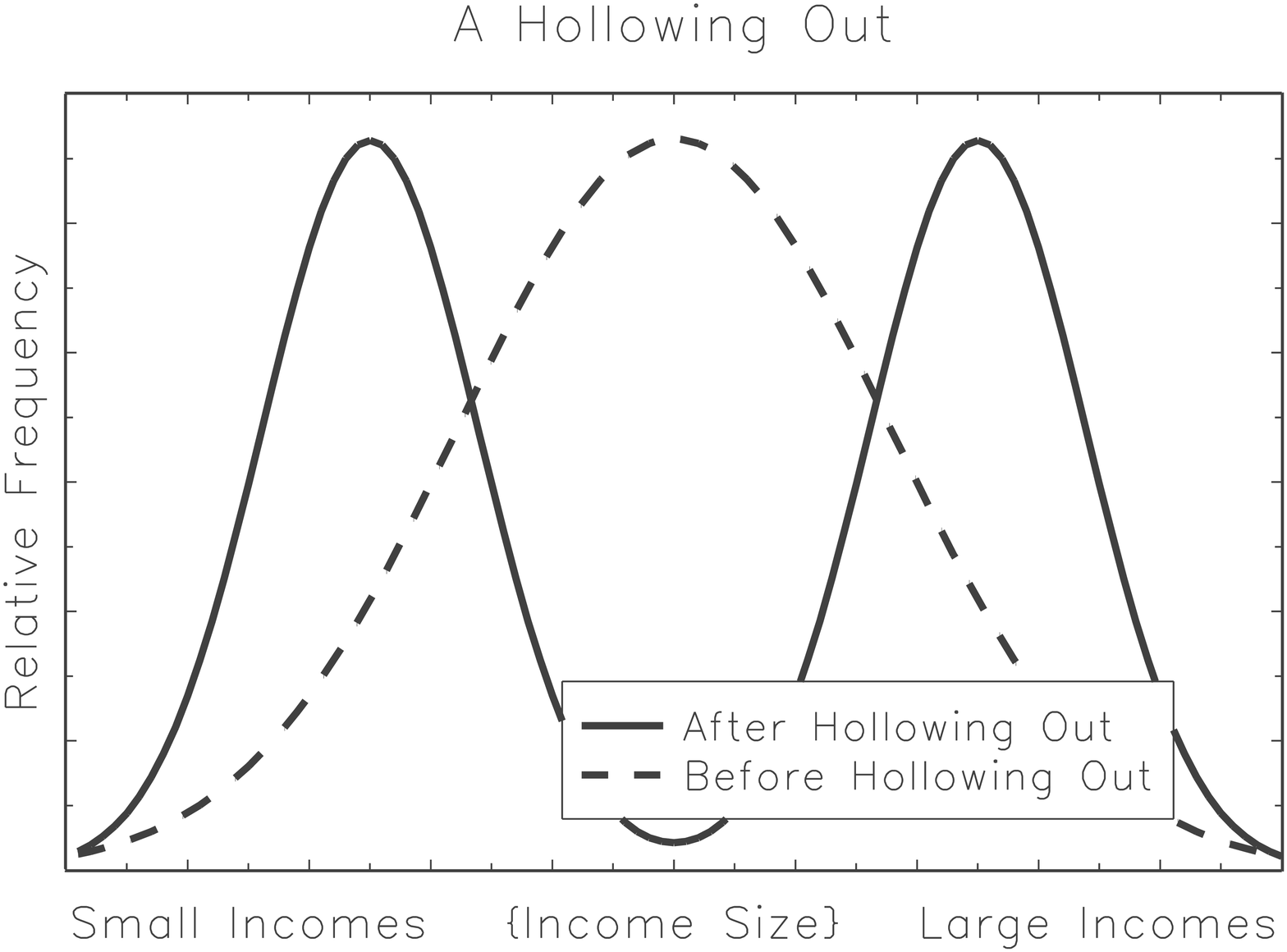}}}
    {\resizebox{6.0cm}{!}{\includegraphics{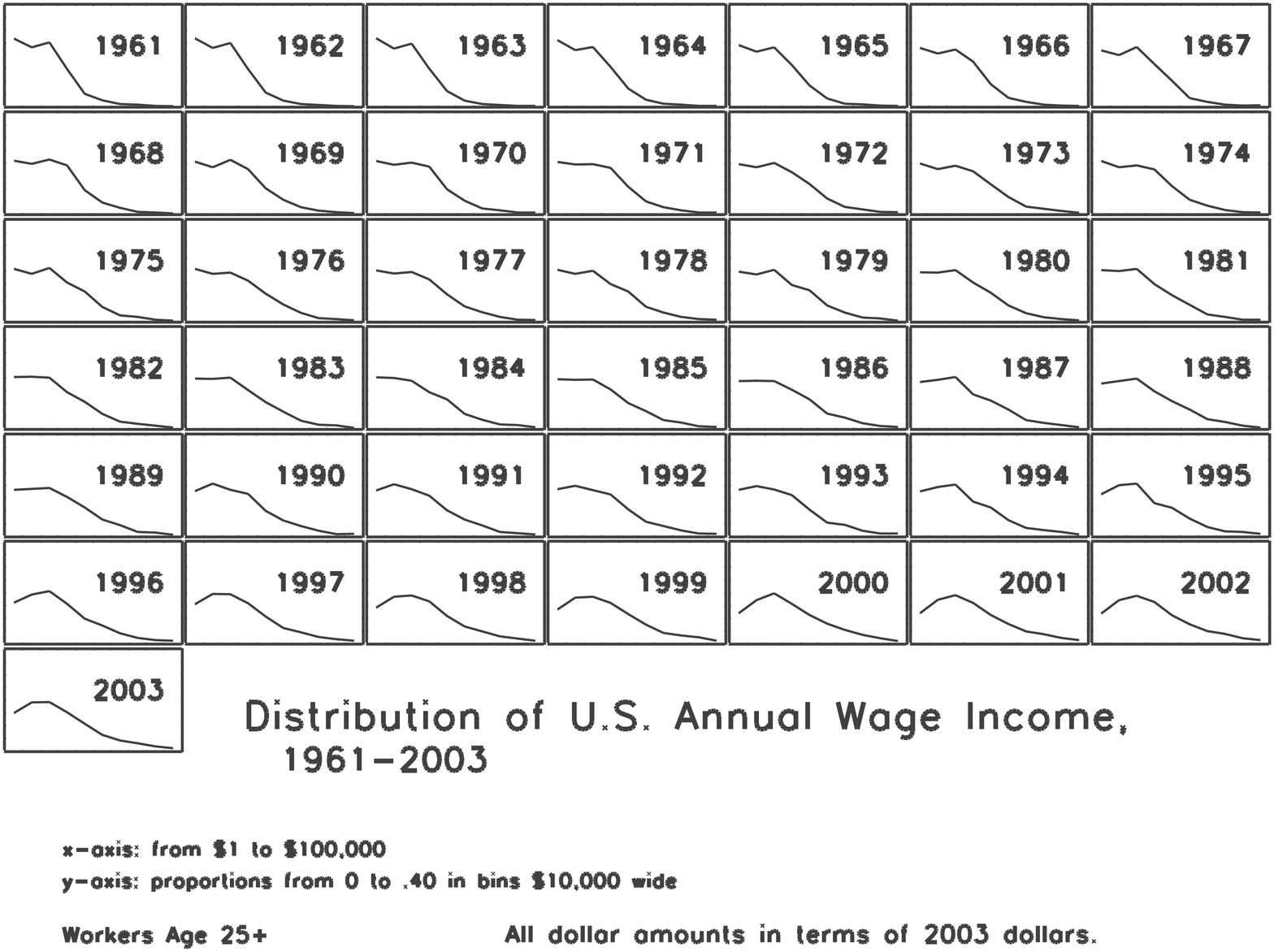}}}
\\    Fig. 3 \hskip 2in Fig. 4
\caption{\label{jafig:fig3}
Caricature of the interpretation in
the popular press of the transformation of the
U.S. wage income distribution into a
bimodal, `barbell' distribution in recent
decades. See Appendix B.
}
\caption{\label{jafig:fig4}
Source: Author's estimates from March CPS data.
}
\end{figure}

The data on which this paper is based are the pooled cross-sectional 
time series formed from `public use samples' of the records 
of individual respondents
to the March Current Population Surveys (CPS) 1962-2004. 
The March CPS is a survey of a large sample of
U.S. households conducted by the U.S. Bureau of the
Census.\footnote{These data have been cleaned, documented 
and made readily accessible as a user-friendly database by Unicon
Research Corporation (2004), a public data reseller supported by 
the (U.S.) National Institutes of Health. The author welcomes
replication of tests of the Inequality Process either with the data 
used in this paper, available at nominal cost from the Unicon
Research Corporation, 111 University Drive, East, Suite 209, 
College Station, Texas 77840, USA, or replication with
comparable data from additional countries.} 
Each March CPS asks questions about the level
of education of members of the household and their
sources of income in the previous calendar year. The
study population is U.S.  residents with at least \$1
(one U.S. dollar) in annual
wage income who are at least 25 years old. All dollar amounts in this
paper are in terms of constant 2003 dollars, i.e., adjusted for changes in
the purchasing power of a dollar from 1961 through 2003. See Appendix A.

Figure 1 illustrates one of two related ways to measure the 
surge in wage income nouveaux riches in the U.S.
1961-2003. Figure 1 measures change in the relative frequencies 
in the far right tail of the wage income distribution, the
distribution of wage income recipients over the largest wage 
incomes. This dynamic of the distribution can be readily
understood in terms of the algebra of the macro model of the 
Inequality Process. The other way to measure the surge is via
change in percentiles of wage incomes (in constant dollars), 
e.g., the 90th percentile. Figure 2 shows the 90th percentile of
wage incomes increasing more in constant dollars than the 10th 
percentile, a small wage income. This second way of
measuring the surge, as a stretching of the distribution to the right, is
implied by the IP's macro model but requires numerical integration to
demonstrate.

The surge in wage income \underline{nouveaux riches} in the U.S.
1961-2003 has been a focus of concern in U.S. labor economics and
sociology journals. A substantial fraction of contributions to this
literature have interpreted the surge as part of the transformation of the
U.S. wage income into a bimodal, U-shaped distribution. 
See figure 3 for a caricature of this thesis and the `hollowing out' literature itself
(Kuttner, 1983; Thurow, 1984; Lawrence, 1984; Blackburn and Bloom, 1985; Bradbury,
1986; Horrigan and Haugen, 1988; Coder, Rainwater, and Smeeding, 1989; Levy and
Michel, 1991; Duncan, Smeeding and Rodgers, 1993; Morris, Bernhardt, and Handcock,
1994; Wolfson, 1994; Esteban and Ray, 1994; Jenkins, 1995; Beach, Chaykowski, and
Slotsve, 1997; Wolfson, 1997; Burkhauser, Crews Cutts, Daly, and Jenkins, 1999;
Esteban and Ray, 1999; Duclos, Esteban and Ray, 2004;).
The emergence of a U-shaped wage income distribution
has been termed the `hollowing out'  or `polarization' of the wage income
distribution. A `hollowed out' distribution has also been called a 
`barbell distribution'. In a `hollowing out' the relative
frequency of middling wage incomes decreases while the relative 
frequencies of small and large wage incomes increase. The
`hollowing out' thesis explains the surge in large wage incomes 
but it is burdened with having to hypothesize a surge in small
wage incomes as well.
The U.S. labor economics and sociology literatures on wage income 
measure trends in terms of scalar statistics of wage
income, mostly the median plus the grab bag of statistics referred 
to under the rubric `statistics of inequality'. Models of the
dynamics of the distribution are rare and never prominent in this 
literature. Thus, hypothesized dynamics of the distribution
have been used in this literature to explain trends in the scalar 
statistics without confirmation of what the empirical distribution
has actually been doing. The rise of the thesis of the `hollowing out' 
of the U.S. wage income distribution requires either that
researchers were unaware of how the empirical distribution of wage 
incomes in the U.S. had changed, or, once the `hollowing
out' interpretation of how the distribution had changed had become 
established in the literature and popularized in the press,
editors and reviewers were unable to accept evidence to the contrary, 
i.e., the journals that established the `hollowing out'
thesis could not correct their error.\footnote{The author tried 
to correct this literature in terms familiar to 
its contributors over a period of many years but, so far, has been unable 
to publish in any of the journals responsible for popularizing the 
`hollowing out' thesis.}

Figure 4 displays estimates of the U.S. wage income distribution 
from 1961 through 2003. It is clear that, indeed, its right
tail thickened over these 43 years, i.e., the relative frequency 
of large wage incomes increased validating half of the `hollowing
out' hypothesis. However, it is as clear in figure 4 that the left 
tail of the distribution of wage incomes, the distribution of
workers over small wage incomes, thinned, that is, the relative 
frequency of small wage incomes decreased. The `hollowing
out' thesis requires both to increase simultaneously.

Figure 5 displays the forward differences between mean relative 
frequencies of wage income between two ten year periods,
1961-1970 and 1994-2003. Ten year means are taken to smooth the 
relative frequencies. Figure 5 shows that the relative
frequency of small wage incomes fell between these two periods 
and larger wage incomes increased, just as an inspection of
figure 4 would lead one to believe.

\section{The Micro- and Macro-Models of the Inequality Process (IP)}

The Inequality Process (IP) is a stochastic binary interacting particle
system derived from an old verbal theory of economic anthropology
(Angle, 1983 to 2006). People are the particles. Wealth is the positive
quantity transferred between particles. The theory from which the IP is
derived is the `surplus theory of social stratification'. It asserts that
economic inequality among people originates in competition for surplus,
societal net product. The IP literature dates from (Angle, 1983). The IP
is an abstraction of a mathematical model from Gerhard Lenski's (1966)
speculative extension of the surplus theory to account for the decrease in
wealth inequality with techno-cultural evolution. Lenski thought that
more skilled workers could bargain for a larger share of what they
produce. Lux (2005) introduced econophysicists to the IP at
Econophys-Kolkata I and pointed out that the econophysics literature on
wealth and income distribution had replicated some of Angle's earlier
Inequality Process findings.

The pair of transition equations of the Inequality Process (IP) for 
competition for wealth between two particles is called
here the micro model of the IP to distinguish the IP's model of 
particle interactions, the micro-level, from the model that
approximates the solution of the micro model in terms of its 
parameters, the distribution of wealth, the macro model.
\begin{figure}
\centering
    {\resizebox{6.0cm}{!}{\includegraphics{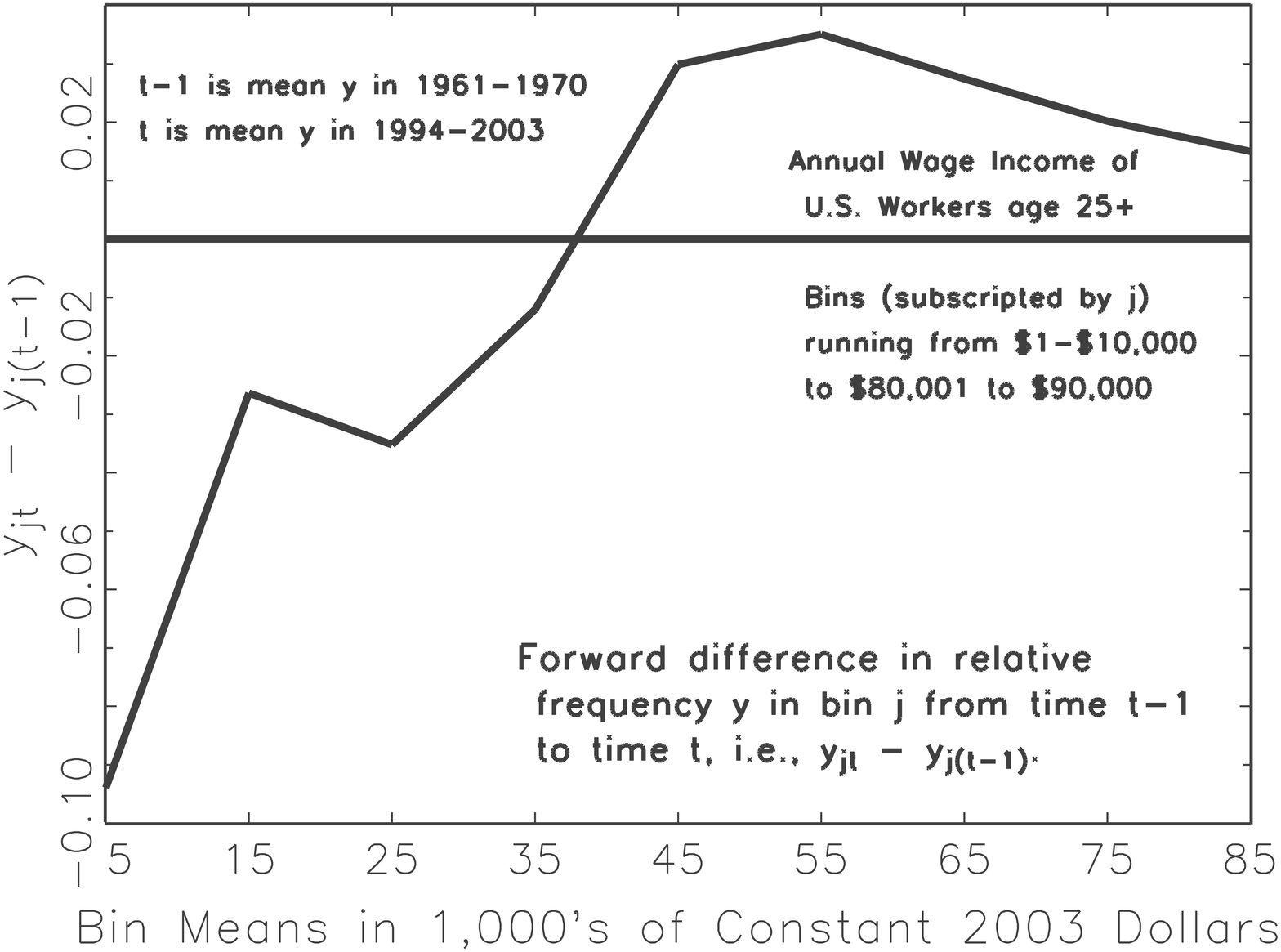}}}
    {\resizebox{6.0cm}{!}{\includegraphics{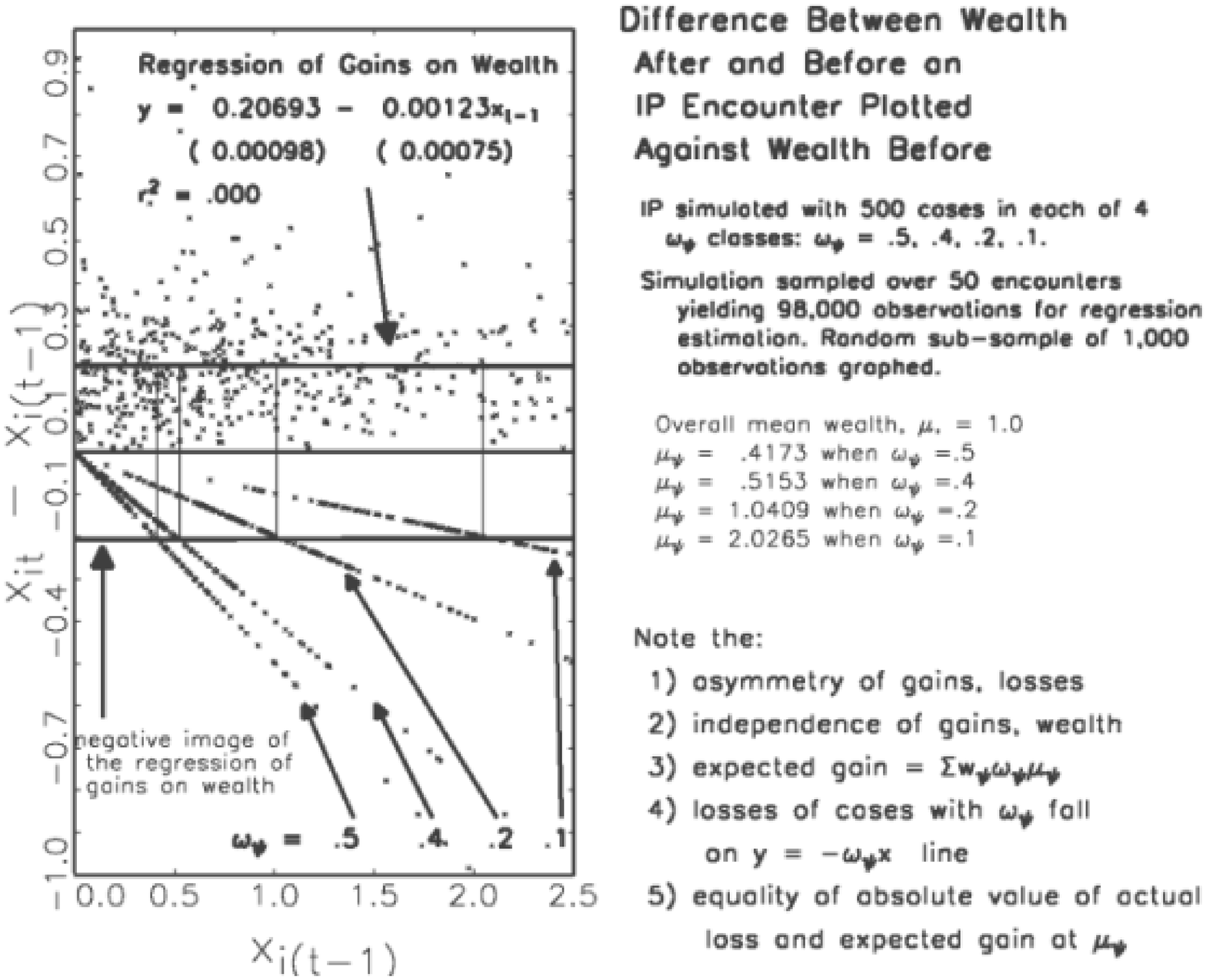}}}
\\    Fig. 5 \hskip 2in Fig. 6
\caption{\label{jafig:fig5}
Forward difference between mean relative
frequencies in 1961-1965 $(t-1)$ and mean relative
frequencies in 1999-2003 $(t)$.
Source: Author's estimates from data of the March CPS.
}
\caption{\label{jafig:fig6}
The scattergram of wealth changes in the population of particles 
from time $t-1$ to $t$ plotted against wealth at time $t-1$.
}
\end{figure}

\subsection{The Micro-Model of the Inequality Process (IP)}
Since the macro-model of the Inequality Process (IP) is derived 
from the IP's micro model, description of the former
should begin with description of the latter. Consider a population 
of particles in which particles have two characteristics.
Wealth is one such trait. Particle $i$'s wealth at time $t$ is denoted, 
$x_{it}$. Wealth is a positive quantity that particles compete for
pairwise in zero-sum fashion, i.e., the sum of wealth of two particles 
after their encounter equals the sum before their
encounter. The other particle characteristic is the proportion of 
wealth each loses when it loses an encounter. That proportion is
the particle's parameter $u$. So, from the point of view of 
a particular particle, say particle $i$, the proportion of wealth it loses, if
it loses, $\omega_i$, is predetermined. When it wins, what it wins is, 
from its point of view, a random amount of wealth. Thus there is
an asymmetry between gain and loss from the point of view of particle $i$. 
Long term each particle wins nearly 50\% of its
encounters. Wealth is transferred, long term, to particles that 
lose less when they lose, the robust losers.

Let particle $i$ be in the class of particles that loses an $\omega_\psi$ 
fraction of their wealth when they lose, i.e., $\omega_i = \omega_\psi$. In the
IP's meta-theory, Lenski's speculation, workers who are more skilled 
retain a larger proportion of the wealth they produce. So
smaller $\omega_\psi$ in the IP's meta-theory represents the more skilled worker. 
Worker skill is operationalized in tests of the IP, as is
usual in labor economics, by the worker's level of education, 
a characteristic readily measured in surveys. For tests of the IP on
wage income data by level of worker education, the IP's population 
of particles is partitioned into equivalence classes of its
particles' $\omega_\psi$ by the corresponding level of education, so that 
the proportion formed by workers at the $\psi$th level of education of
the whole labor force, $u_\psi$, (`$u$' to suggest `weight'), is 
the proportion formed by the $\omega_\psi$ equivalence class of the 
population of particles. The $u_\psi$'s are estimated from data and so are 
the $\omega_\psi$'s by fitting the comparable statistic of the IP to 
either micro-level data (the dynamics of individual wage incomes) 
or macro-level data (the dynamics of wage income distributions). The IP's
meta-theory implies that estimated 
$\omega_\psi$'s should scale inversely with worker 
education level, based on the assumption that the more educated worker 
is the more productive worker. Nothing in the testing of this 
prediction forces this outcome. The
predicted outcome holds in U.S. data on wage incomes by level 
of education as demonstrated in Angle (2006) for 1961-2001
and below for 1961-2003. While there is no apparent reason why 
this finding should not generalize to all industrial labor
forces in market economies, the universality of the prediction 
is not yet established. The IP is a highly constrained model. Its
predictions are readily falsified if not descriptive of the data.

The transition equations of the competitive encounter for wealth 
between two particles in the Inequality Process' (IP's)
micro model are:
\begin{eqnarray}
\label{ja:micipeq}
x_{it} &=& x_{i(t-1)} + \omega_{\theta j} d_{it} x_{j(t-1)}
- \omega_{\psi i} (1-d_{it}) x_{i(t-1)}\nonumber\\
x_{jt} &=& x_{j(t-1)} - \omega_{\theta j} d_{it} x_{j(t-1)}
+ \omega_{\psi i} (1-d_{it}) x_{i(t-1)}
\end{eqnarray}
where $x_{i(t-1)}$ is particle $i$'s wealth at time $t-1$,
\[
d_{it}=\left\{ \begin{array}{cl}
1 & \ {\rm with \ probability} \ .5 \ {\rm at \ time} \ t\\
0 & \ {\rm otherwise.}
\end{array}\right.
\]
and,

$\omega_{\psi i}$ = proportion of wealth lost by particle $i$ 
when it loses (the subscript indicates that particle $i$ has a parameter whose
value is $\omega_{\psi i}$; there is no implication that the $\omega_\psi$ 
equivalence class of particles has only one member or that
necessarily $\omega_{\psi i} \ne \omega_{\theta j}$);

$\omega_{\theta j}$ = proportion of wealth lost by particle $j$ when it loses.

Particles are randomly paired; a winner is chosen via a discrete 0,1 
uniform random variable; the loser gives up a fixed
proportion of its wealth to the winner. In words, the process is:

Randomly pair particles. One of these pairs is particle $i$ and particle $j$.
A fair coin is tossed and called. If particle $i$
wins, it receives an $\omega_\theta$ share of particle $j$'s wealth. 
If particle $j$ wins, it receives an $\omega_\psi$ share of particle 
$i$'s wealth. The other particle encounters are analogous. Repeat.

The asymmetry of gain and loss is apparent in figure 6, the graph of 
forward differences, $x_{it} - x_{i(t-1)}$ against wealth, $x_{i(t-1)}$,
resulting from (\ref{ja:micipeq}).
The Inequality Process differs
from the Saved Wealth Model, a modification of the stochastic model of 
the Kinetic Theory of Gases that generates a gammoidal stationary distribution 
discussed by Chakraborti, Chakrabarti
(2000); Chatterjee, Chakrabarti, and Manna (2003); 
Patriarca, Chakraborti, and Kaski (2004);
Chatterjee, Chakrabarti, and Manna (2004); 
Chatterjee, Chakrabarti, and Stinchcombe (2005);
Chatterjee, Chakraborti, and Stinchcombe (2005); 
Patriarca, Chakraborti, Kimmo, and Germano
(2005). The following substitution converts the 
Inequality Process into the Saved Wealth Model
(apart from the random $\omega$ factor in Chatterjee et al, 2004
and subsequent papers):
\[ d_{it} \to \epsilon_{it} \]
where $\epsilon_{it}$ is a continuous, uniform i.i.d 
random variate with support at $[0.0,1.0]$.

\subsection{The Macro-Model of the Inequality Process (IP)}
The macro model of the Inequality Process (IP) is a gamma 
probability density function (pdf), $f_{\psi t}(x)$, a model of the 
wage income, $x$, of workers at the same level of education,
the $\psi$th at time $t$. The macro model approximates the 
stationary distribution of wealth of the IP's
micro model. The macro model was developed in a chain of papers 
(Angle, 1993, 1996-2001, 2002b-2006). 
The IP's macro model in the $\omega_\psi$ equivalence class is:
\begin{equation}
\label{ja:fpsit}
f_{\psi t}(x)= \frac{\lambda_{\psi t}^{\alpha_\psi}} {\Gamma (\alpha_\psi)}
x^{\alpha_\psi -1} \exp(-\lambda_{\psi t} x)
\end{equation}
or in terms of the IP's parameter in the $\omega_\psi$ equivalence class:
\begin{eqnarray}
\label{ja:fpsit1}
f_{\psi t}(x) &=& \exp \left[ 
\left( \frac{1-\omega_\psi}{\omega_\psi}\right)
\ln \left( \frac{1-\omega_i}{\tilde{\omega}_t \mu_t} \right) \right] \nonumber \\
&& \times \exp \left[ - \ln \Gamma \left( \frac{1-\omega_\psi}{\omega_\psi} \right)
+ \left( \frac{1-2\omega_i}{\omega_i} \right)\ln (x)
- \left( \frac{1-\omega_i}{\tilde{\omega_t} \mu_t} \right) x
\right]
\end{eqnarray}
where:

$\alpha_\psi \equiv$ the shape parameter of the gamma pdf that 
approximates the distribution of wealth, $x$, in
the $\omega_\psi$ equivalence class, intended to model the 
wage income distribution of workers at the $\psi$th level of 
education regardless of time; $\alpha_\psi > 0$
\begin{equation}
\label{ja:alphapsiequiv}
\alpha_\psi \approx \frac{1-\omega_\psi}{\omega_\psi}
\end{equation}
and:

$\lambda_{\psi t} \equiv$ scale parameter of distribution of the 
gamma pdf that approximates the distribution
of wealth, $x$, in the $\omega_\psi$ equivalence class, 
intended to model the wage income distribution of workers at level $\psi$
of education in a labor force with a given unconditional mean of 
wage income and a given harmonic mean of $\omega_\psi$'s at time $t$;

$\lambda_{\psi t} > 0$
\begin{equation}
\label{ja:lambdapsit1}
\lambda_{\psi t} \approx \frac{(1-\omega_\psi)\left( \frac{u_{1t}}{\omega_1}
+ \ldots + \frac{u_{\psi t}}{\omega_\psi} 
+ \ldots + \frac{u_{\Psi t}}{\omega_\Psi}\right)}
{\mu_t} \approx \frac{(1-\omega_\psi)}{\tilde{\omega_t} \mu_t}
\end{equation}
where:

$\mu_t =$ unconditional mean of wage income at time $t$

$\tilde{\omega_t} =$ harmonic mean of the $\omega_\psi$'s at time $t$.

\noindent
and $\mu_t$ and the $u_{\psi t}$'s are exogenous and the sole source of 
change in a population of particles where $\Psi$ $\omega$ equivalence 
classes are distinguished. Consequently, the dynamics of (2), the IP's 
macro model, are exogenous, that is, driven by the product 
($\tilde{\omega_t} \mu_t$) and expressed as a scale transformation, i.e.,
via $\lambda_{\psi t}$. Figure 7 shows the shapes of a gamma probability 
density function (pdf) for a fixed scale parameter and several values 
of the shape parameter, $\alpha_\psi$. Figure 7 shows that if the IP's 
meta-theory is correct, more education,
operationalized as smaller $\omega_\psi$ earns a worker a place
in a wage income distribution with a larger $\alpha_\psi$, a more centralized
distribution, whose mean, equal to $\alpha_\psi / \lambda_{\psi t}$, 
is larger than that of the worker with less education.
\begin{figure}
\centering
    {\resizebox{5.5cm}{!}{\includegraphics{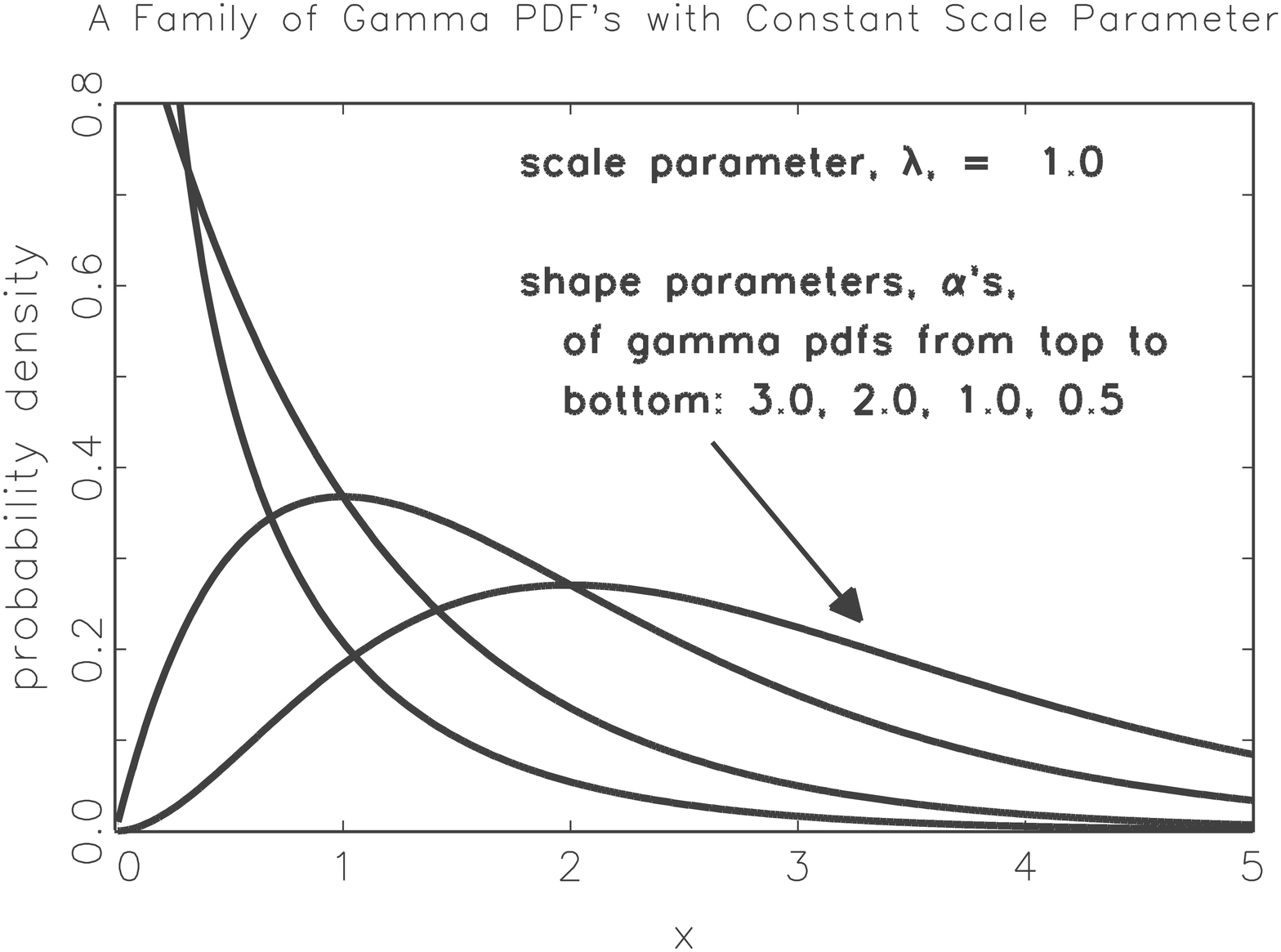}}}
    {\resizebox{5.5cm}{!}{\includegraphics{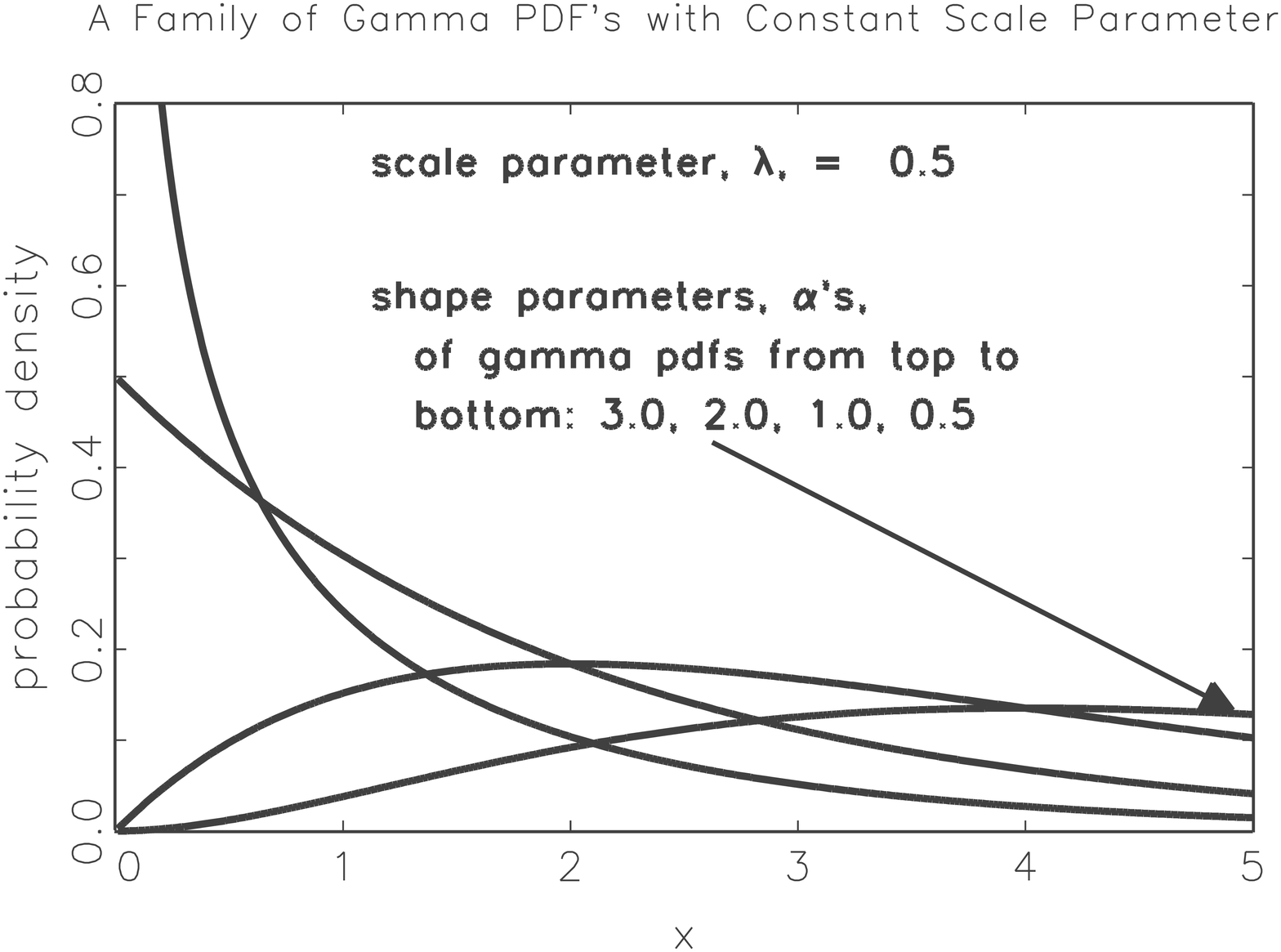}}}
\\    Fig. 7 \hskip 2in Fig. 8
\caption{\label{jafig:fig7}
A family of gamma pdfs with different
shape parameters but the same scale parameter, $1.0$.
}
\caption{\label{jafig:fig8}
A family of gamma pdfs with different
shape parameters but the same scale parameter, $0.5$.
}
\end{figure}
\begin{figure}
\centering
    {\resizebox{6.0cm}{!}{\includegraphics{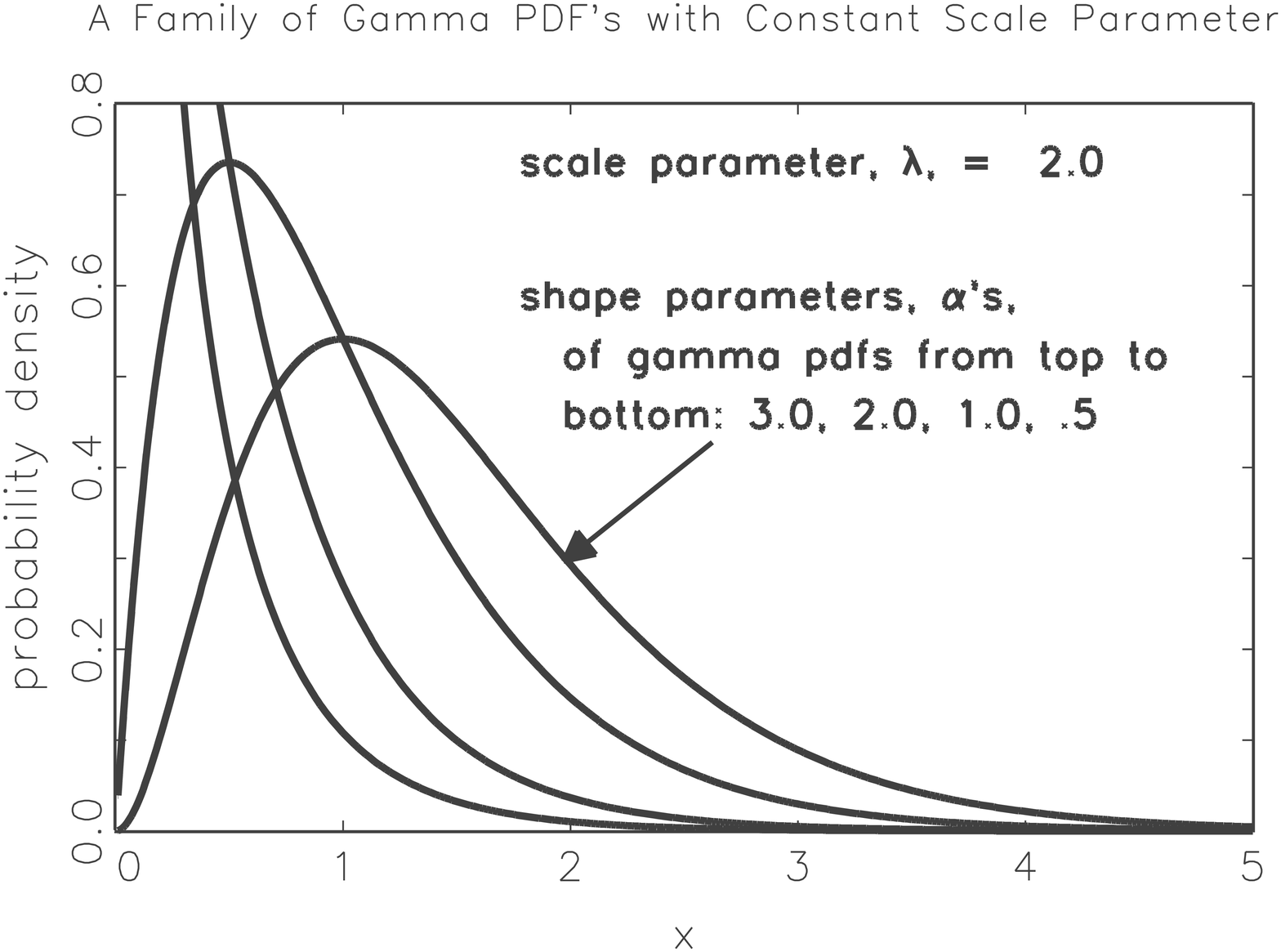}}}
    {\resizebox{6.0cm}{!}{\includegraphics{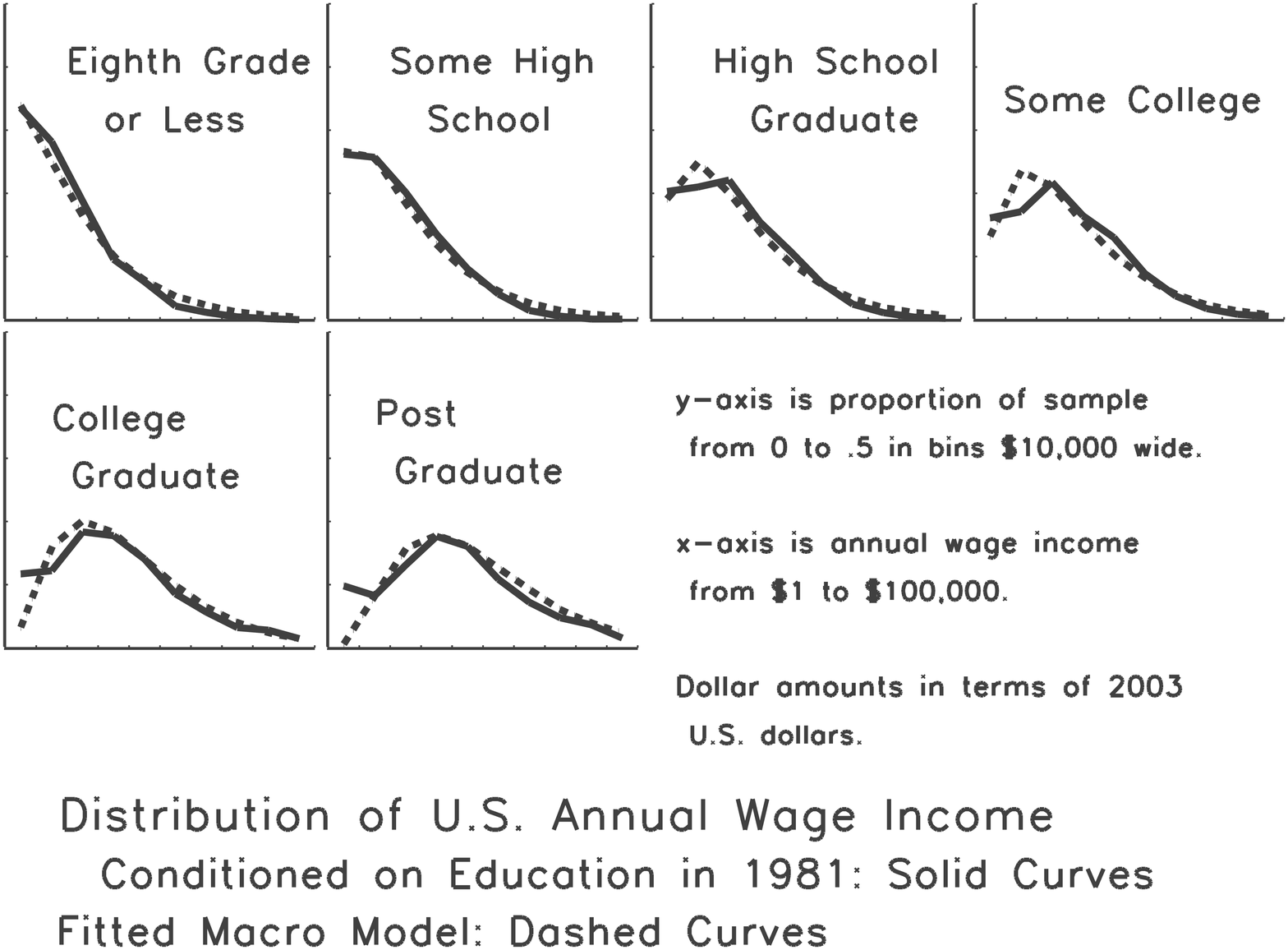}}}
\\    Fig. 9 \hskip 2in Fig. 10
\caption{\label{jafig:fig9}
A family of gamma pdfs with different
shape parameters but the same scale parameter, $2.0$.
}
\caption{\label{jafig:fig10}
Source: author's estimates based on March, CPS data.
}
\end{figure}

Comparison of figures 7, 8, and 9 show the consequences of change in
the gamma scale parameter on a gamma distribution holding the shape
parameters constant. A decrease in $\lambda_\psi$ stretches the mass 
of the pdf to the right over larger $x$'s as in figure 8, 
increasing all percentiles of $x$. Compare figure 8
to figure 7. In the IP, particles circulate randomly within the 
distribution of wealth of their $\omega_\psi$ equivalence class, 
so an increase in $\mu_t$ of a magnitude
sufficient to increase the product ($\tilde{\omega_t} \mu_t$) 
and decrease $\lambda_{\psi t}$ may not mean that
each and every particle in the $\omega_\psi$ equivalence class increases its wealth,
although all the percentiles increase. $\tilde{\omega_t}$ is expected 
to decrease given a rising level of education in the U.S. labor force. 
$\mu_t$, the unconditional mean of wage income, rose irregularly 
in the U.S. in the last four decades of the 20th century.

If proportional increase in $\mu_t$ offsets proportional decrease
in $\tilde{\omega_t}$ then the product ($\tilde{\omega_t} \mu_t$) increases, 
$\lambda_{\psi t}$ decreases, and the IP's macro model
implies that wage income distribution is stretched 
to the right as in figure 8 with
all percentiles of wage income increasing. However, if the product 
($\tilde{\omega_t} \mu_t$) decreases, then $\lambda_{\psi t}$ increases 
and the IP's macro model predicts that the wage
income distribution is compressed to the left, that is, its mass is moved over
smaller wage income amounts and its percentiles decrease as in figure 9 by
comparison to figures 7 and 8.

The product ($\tilde{\omega_t} \mu_t$) is estimated in the fit of the 
IP's macro model to the 43 distributions of wage income conditioned 
on education in the U.S. according to the March CPS' of 1962 
through 2004, which collected data on wage incomes
in 1961 through 2003. Six levels of worker education level have been
distinguished. See Table 1. There are 43 X 6 = 258 partial 
distributions to be fitted. Also fitted are 258 median wage incomes,
one for each partial distribution fitted. See Appendix B. 
Each partial distribution has fifteen relative frequency bins, each
\$10,000 (in constant 2003 dollars) wide, e.g., \$1 - \$10,000, 
\$10,001-\$20,000, etc, for a total of 258 X 15 = 3,870 $x$ (income),
$y$ (relative frequency) pairs to be fitted by the IP's macro 
model which has six degrees of freedom, the six values of $\omega_\psi$
estimated. The fits are simultaneous. The fitting criterion is the 
minimization of weighted squared error, i.e., nonlinear least
squares. The weight on each partial distribution in each year in the 
fit is $u_{\psi t}$, the proportion of the labor force with $\psi$th level of
education in that year. A search is conducted over the parameter 
vector via a stochastic search algorithm that is a variant of
simulated annealing to find the six values that minimize squared error. 
The squared correlation between the 3,870 observed
and expected relative frequencies is .917. Table 1 displays the estimated 
parameters and their bootstrapped standard errors.
Note that the estimated $\omega_\psi$'s scale inversely with level 
of education as predicted by the IP's meta-theory. Figure 10 displays the
IP macro model's fit to the six partial distributions of wage 
income by level of education in 1981.
\begin{table}
\caption{Estimates of the Parameters of the IP's Macro-Model}
\label{ja:table1}
\begin{tabular}{|l|l|l|l|}
\hline
Highest Level &
$\omega_\psi$ estimated by fitting &
bootstrapped &
estimate of $\alpha_\psi$ \\

of Education & 
the macro-model to 258 &
standard error &
corresponding \\

 & partial distributions &
of $\omega_\psi$ & to $\omega_\psi$  \\

 & (43 years X 6 levels &  (100 re-samples) & \\

& of education) & & \\
\hline
eighth grade or less & 0.4524 & .0009582 & 1.1776\\

\hline
some high school & 0.4030 & .0006159 & 1.4544\\

\hline
high school graduate & 0.3573  &.0004075 & 1.7924 \\

\hline
some college & 0.3256 & .0005033 & 2.0619 \\

\hline
college graduate & 0.2542 & .0007031 & 2.7951 \\

\hline
post graduate education & 0.2084 & .0005216 & 3.6318 \\

\hline
\end{tabular}
\end{table}

Separately, 258 gamma pdfs, each with two unconstrained parameters,
were also fitted to each of the 258 partial distributions, 
a 516 parameter fit. These
fits were done to create an alternative model to baseline 
how much less well the
IP's macro model did than unconstrained gamma pdf fits to the same data set.
The squared correlation between the 3,870 observed and expected
relative frequencies under this alternative model is .957. Thus the
IP's macro model fits the data almost as well as the unconstrained
gamma pdf alternative model although the IP's macro model uses
only 6 degrees of freedom and the alternative model 516.
\begin{figure}
\centering
    {\resizebox{6.0cm}{!}{\includegraphics{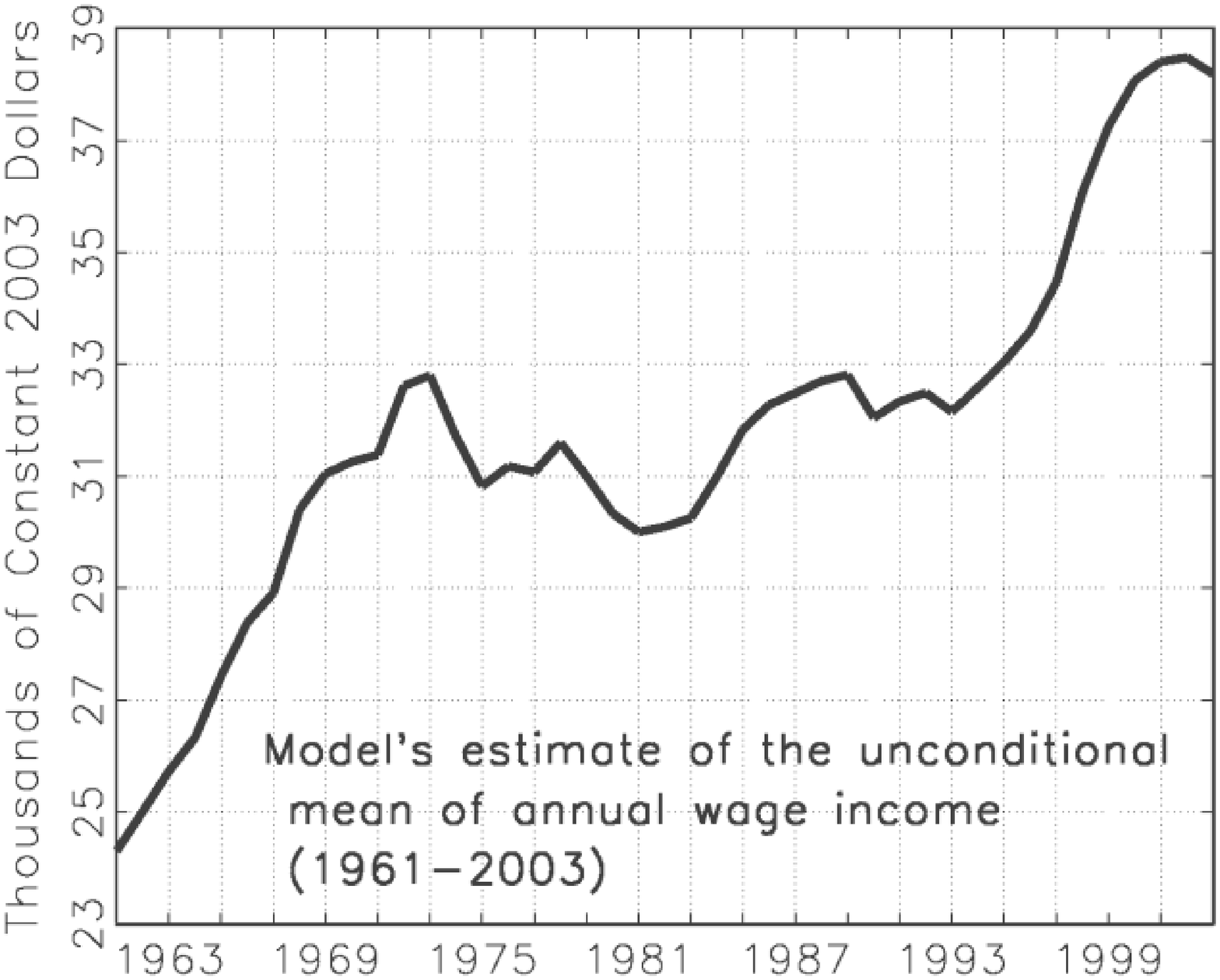}}}
    {\resizebox{6.0cm}{!}{\includegraphics{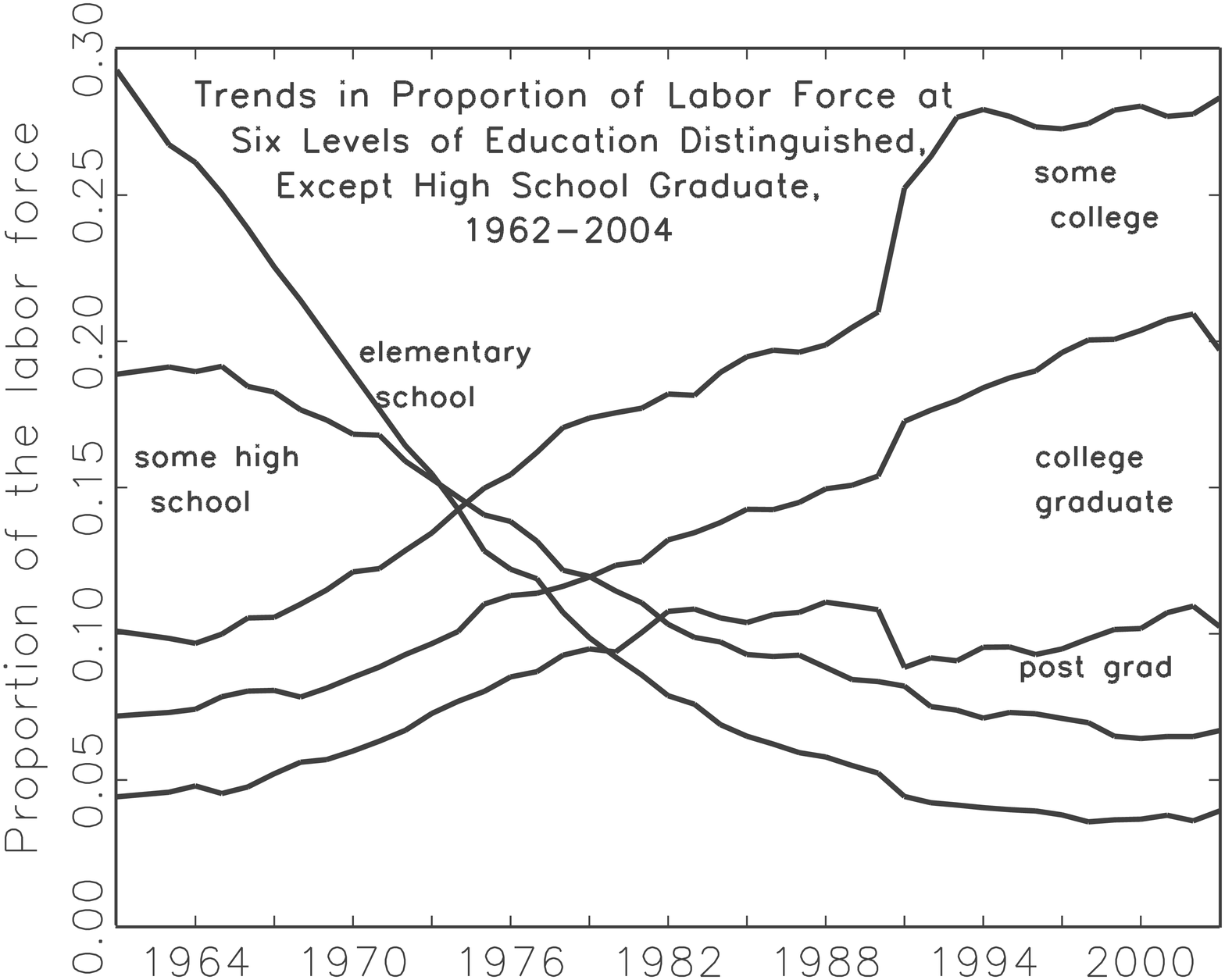}}}
\\    Fig. 11 \hskip 2in Fig. 12
\caption{\label{jafig:fig11}
Source: Author's estimates from data of the March Current Population Survey.
}
\caption{\label{jafig:fig12}
Source: Author's estimates from data of the March Current Population Survey.
}
\end{figure}

Figure 11 shows that the unconditional mean of wage income in the U.S.
increased substantially in the 1960's and again in the 1990's in constant 2003
dollars. There was a smaller move upward in the early to mid 1980's.
However between the early 1970's and mid-1990's there were small
declines and small increases that netted each other out, 
i.e., the unconditional mean of wage income in the U.S. did not 
increase in constant dollar terms for over two decades. The IP's 
macro model implies that the scale factor of wage income at each level of 
education in the labor force is driven by the product 
($\tilde{\omega_t} \mu_t$). In the model bigger ($\tilde{\omega_t} \mu_t$)
stretches the distribution of wage incomes at each
level of education to the right over larger wage incomes. $\mu_t$ 
has to increase proportionally more than $\tilde{\omega_t}$ 
decreases for all percentiles of the distribution
conditioned on education to increase.
\begin{figure}
\centering
    {\resizebox{6.0cm}{!}{\includegraphics{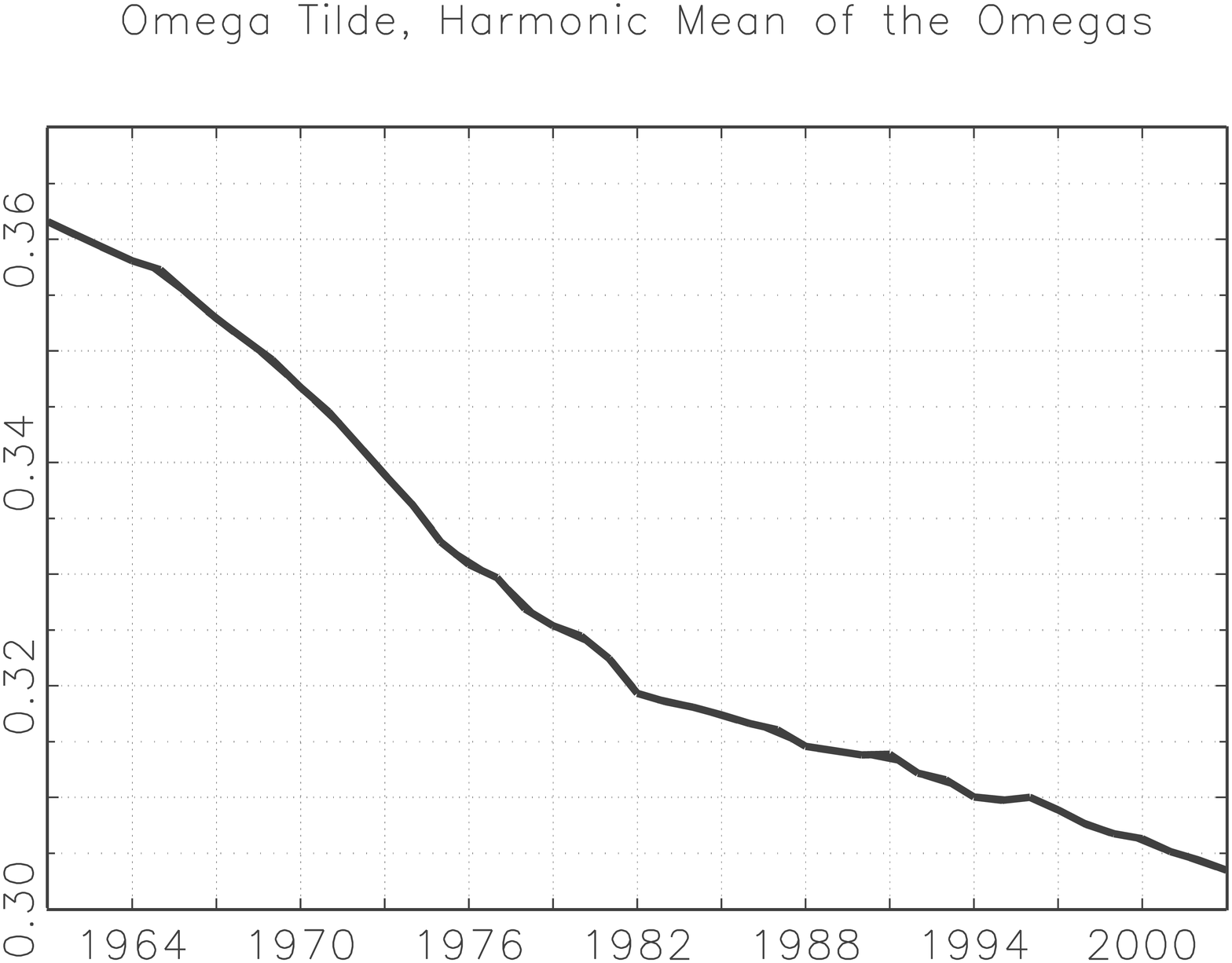}}}
    {\resizebox{6.0cm}{!}{\includegraphics{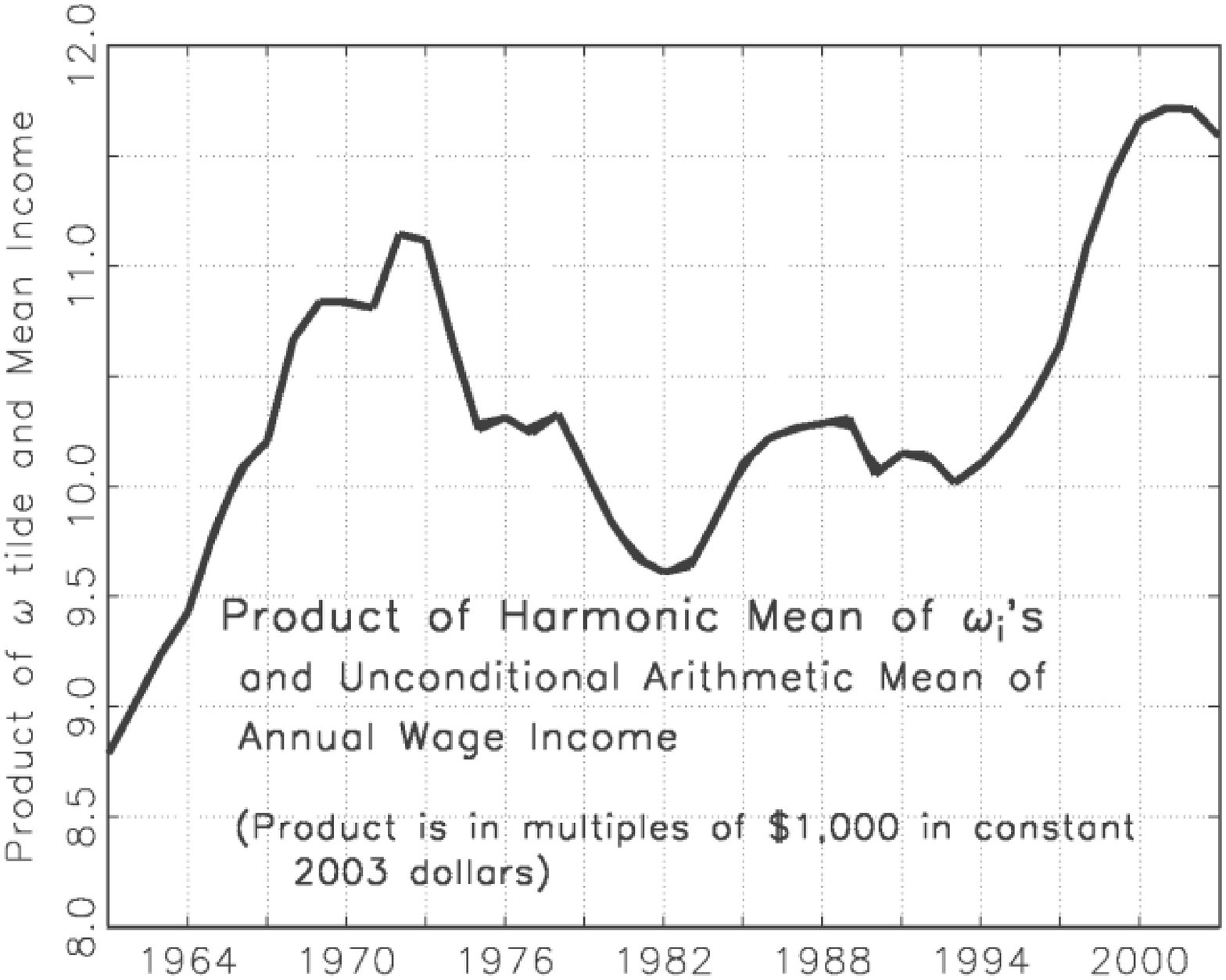}}}
\\    Fig. 13 \hskip 2in Fig. 14
\caption{\label{jafig:fig13}
Source: Author's estimates from data of the March Current Population Survey.
}
\caption{\label{jafig:fig14}
Product of the unconditional mean, $\mu_t$,
and the harmonic mean of the $\omega_\psi$'s, $\tilde{\omega}_t$.
Source: Author's estimates from data of the March Current Population Survey.
}
\end{figure}

$\tilde{\omega_t}$ is the harmonic mean of the $\omega_\psi$'s 
at each time point. The proportion each $\omega_\psi$ 
equivalence class forms of the population, $u_{\psi t}$, 
changes as the proportion of workers at a given educational level changes 
in the labor force. From 1961-2003 the level of education of the U.S. 
labor force rose substantially. See figure 12. Given the $\omega_\psi$'s 
estimates in Table 1, $\tilde{\omega_t}$ decreases 1961-2003 as 
the level of education rises in the U.S. labor force. Figure 13 displays 
the course of $\tilde{\omega_t}$ from 1961 through 2003, 
a steady decline throughout.
\begin{figure}
\centering
    {\resizebox{6.0cm}{!}{\includegraphics{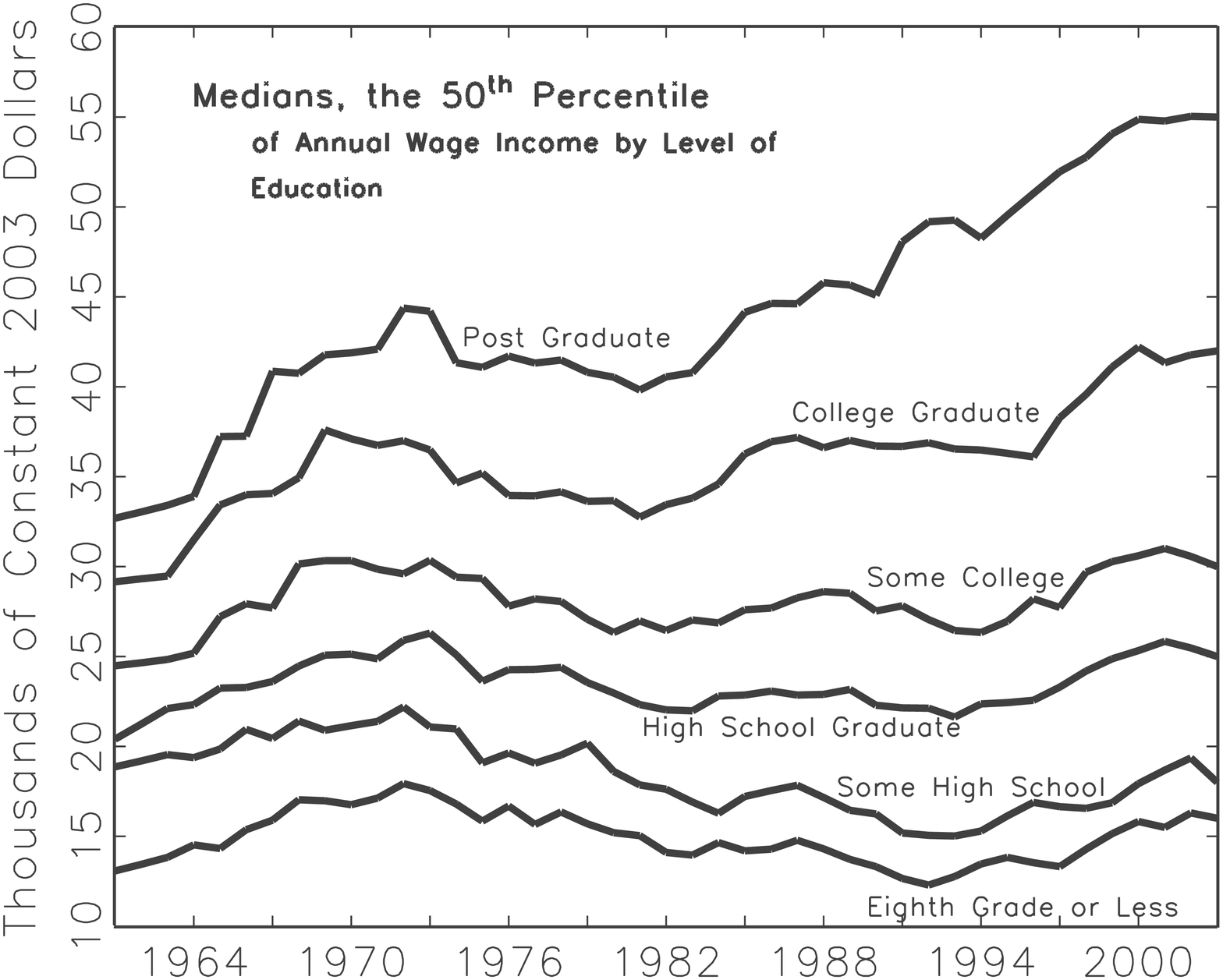}}}
    {\resizebox{6.0cm}{!}{\includegraphics{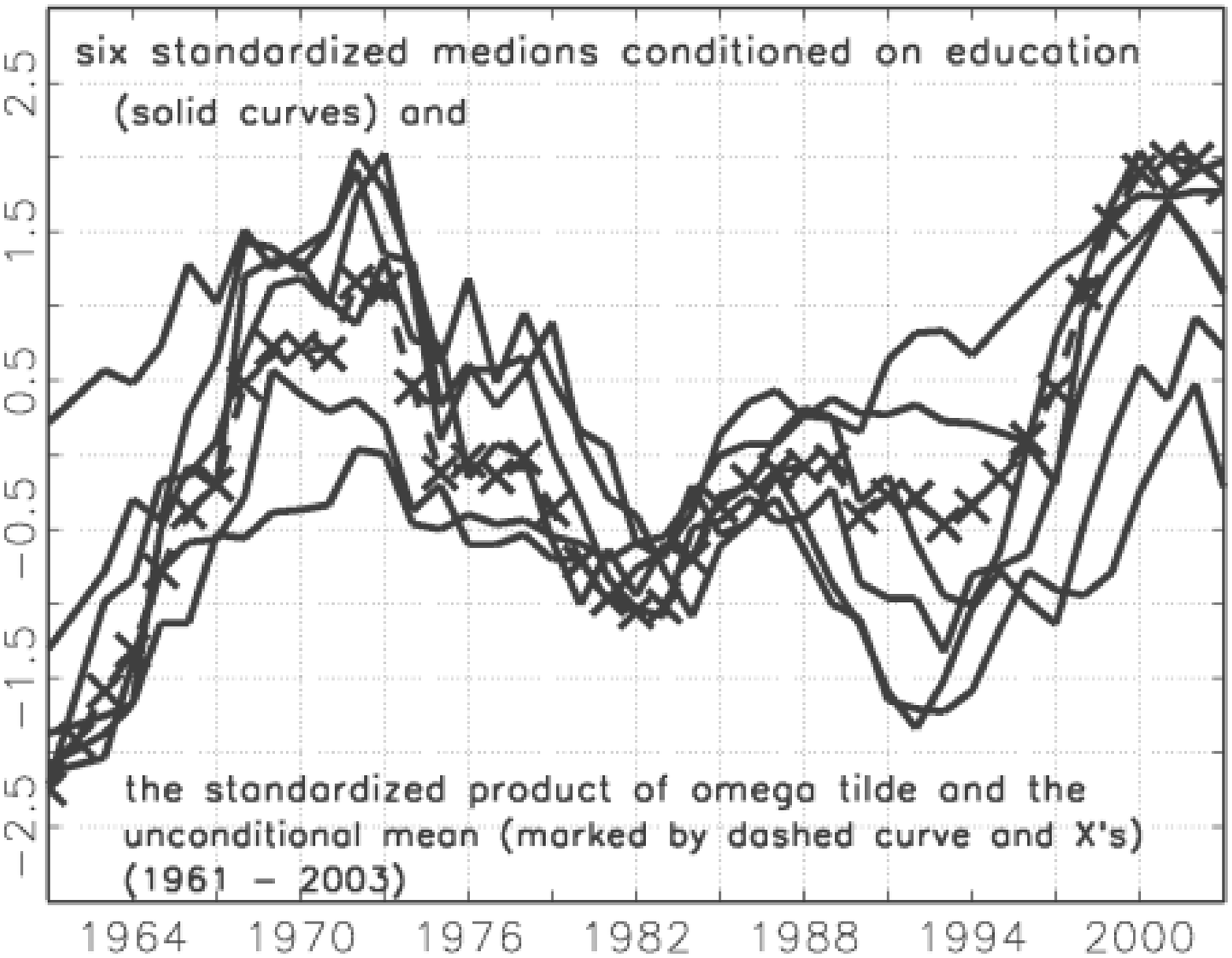}}}
\\    Fig. 15 \hskip 2in Fig. 16
\caption{\label{jafig:fig15}
Source: Author's estimates from data of the March Current Population Survey.
}
\caption{\label{jafig:fig16}
Source: Author's estimates from data of the March Current Population Survey.
}
\end{figure}

Figure 14 graphs the estimated product ($\tilde{\omega_t} \mu_t$) 
over time. Note that decline between the early 1970's and 
mid-1990's was much larger proportionally
than in figure 11, the time-series of $\mu_t$. 
That means that the U.S. wage income distribution conditioned on 
education was compressed substantially to the
left over smaller wage incomes from 1976 through $1983$ in a much more
pronounced way than figure 11 implies. Figure 11 incorporates the 
positive effect on the unconditional mean of the rise in education 
level in the U.S. labor force from 1961 through 2003. Figure 14
shows the negative effect of the rise in educational level on wage earners,
few of whom raise their education level while they work for a
living. The rise in education level in the labor force as a whole is due to
the net increase occasioned by the
entry of more educated younger workers and the exit of less educated older
workers. Figure 14 shows that most wage earners, those not raising their
education level while they worked, experienced a decrease of wage income
percentiles during the 1970's and early 1980's, the sort of wage income
compression toward smaller wage incomes shown in the comparison of 
figure 9 to figure 8.

Figure 15 confirms that
smaller ($\tilde{\omega_t} \mu_t$) did result, as the IP's
macro model implies, in downturns in the medians of wage earners at
each level of education 
from the early 1970's through the 1990's with an exception in the high, 
`open-end' category of education. Its mean level of education rose.
Standardization of the time-series of ($\tilde{\omega_t} \mu_t$) 
in figure 14 and standardization of the 6 time-series of conditional 
median wage incomes in figure 15 allow direct observation of how 
closely these six are associated with ($\tilde{\omega_t} \mu_t$). 
Figure 16 shows the graphs of the 7 standardized time-series. 
The time-series of the standardized ($\tilde{\omega_t} \mu_t$)'s is marked 
by $X$'s. The 6 standardized medians track the standardized 
($\tilde{\omega_t} \mu_t$)'s. Table 2 shows that 4 of the 6 time-series 
of conditional medians are more closely correlated with the product 
($\tilde{\omega_t} \mu_t$) than with the unconditional mean, $\mu_t$, alone.
The IP's macro model implies a larger correlation between the time-series of
median wage income conditioned on education and ($\tilde{\omega_t} \mu_t$) 
than with $\mu_t$ alone. This inference follows from Doodson's
approximation formula for the median of the gamma pdf, 
$f_{\psi t}(x)$, $x_{(50)\psi t}$ (Weatherburn, 1947:15
[cited in Salem and Mount, 1974: 1116]):

Mean - Mode $\approx$ 3 (Mean - Median)

\noindent
in terms of a two parameter gamma pdf:
\[
x_{(50)\psi t} \approx \frac{3 \alpha_\psi -1}{3 \lambda_{\psi t}}
\]
and given (4) and (5):
\begin{equation}
\label{ja:x50psit_a}
x_{(50)\psi t} \approx \left( \frac{1- \frac{4}{3} \omega_\psi}{1-\omega_\psi} \right) \left( \frac{\tilde{\omega_t} \mu_t}{\omega_\psi}\right)
\end{equation}
a constant function of the conditional mean, 
$(\tilde{\omega_t} \mu_t) /\omega_\psi$.

\begin{table}
\caption{Estimated Correlations Between Time-Series}
\label{ja:table2}
\begin{tabular}{|l|l|l|}
\hline
Highest Level of & correlation between $\tilde{\omega_t} \mu_t$
& correlation between the \\
Education & and median wage income & unconditional mean $\mu_t$\\
& at a given level of & and median wage income at \\
& education & a given level of education\\
\hline
eighth grade or less & .5523 & .1837 \\
\hline
some high school & .1573 & -.2874 \\
\hline
high school graduate & .8729 & .5885 \\
\hline
some college & .9279 & .7356 \\
\hline
college graduate & .9042 & .9556 \\
\hline
post graduate education & .7776 & .9575 \\
\hline
\end{tabular}
\end{table}

\section{The Dynamics of the Macro Model of the Inequality Process (IP)}
The dynamics of the Inequality Process (IP)'s macro-model 
of the wage income distribution of workers at the same level of
education are driven exogenously by change in $(\tilde{\omega_t} \mu_t)$:
\begin{eqnarray}
\label{ja:delfpsit}
\frac{\partial f_{\psi t} (x)}{\partial (\tilde{\omega_t} \mu_t)}
&=& f_{\psi t} (x) \ \lambda_{\psi t}
\left( \frac{x-\mu_{\psi t}}{\tilde{\omega_t} \mu_t} \right) \nonumber\\
&=& f_{\psi t} (x) \
\frac{(1-\omega_\psi)}{(\tilde{\omega_t} \mu_t)^2} \
(x-\mu_{\psi t})
\end{eqnarray}
where, the conditional mean of wealth in the $\omega_\psi$
equivalence class, $\mu_t$, is:
\begin{equation}
\label{ja:mupsit}
\mu_{\psi t} = \frac{\alpha_\psi}{\lambda_{\psi t}} \approx \frac{\tilde{\omega_t} \mu_t}{\omega_\psi}
\end{equation}
In (7), as $(\tilde{\omega_t} \mu_t)$ increases, $f_{\psi t}(x_0)$ 
decreases to the left of the conditional mean, $\mu_{\psi t}$, i.e., 
for $x_0 < \mu_{\psi t}$.  $f_{\psi t}(x_0)$ increases to the right of 
the conditional mean $\mu_{\psi t}$, i.e., for 
$x_0 > \mu_{\psi t}$. So an increase in 
$(\tilde{\omega_t} \mu_t)$  simultaneously thins the
left tail of the distribution of $x$, wealth, in the $\omega_\psi$ 
equivalence class and thickens the right tail. 
(7) implies that the probability mass in the left and right tails, 
defined as the probability mass over $x_0 < \mu_{\psi t}$ and 
$x_0 > \mu_{\psi t}$ respectively, must vary inversely if
$(\tilde{\omega_t} \mu_t)$ changes. Thus, the macro-model of the 
Inequality Process (IP) squarely contradicts the hypothesis that a wage
distribution conditioned on education can become U-shaped via a simultaneous 
thickening of the left and right tails, what the
literature on the `hollowing out' of the U.S. distribution of 
wage incomes asserts.

Given that in the IP's macro-model all change is exogenous, due to 
$(\tilde{\omega_t} \mu_t)$, the forward difference, 
$f_{\psi t} (x_0) - f_{\psi (t-1)} (x_0)$, at a given $x_0$ can be 
approximated via Newton's approximation as:
\begin{eqnarray}
\label{ja:fpsitx0}
f_{\psi t}(x_0)-f_{\psi (t-1)} (x_0) &\approx&
f_{\psi (t-1)} (x_0) + f^\prime_{\psi (t-1)} (x_0)
\left( (\tilde{\omega_t} \mu_t) - (\tilde{\omega}_{t-1} \mu_{t-1})\right) \nonumber \\
&& -f_{\psi (t-1)} (x_0) \nonumber \\
&\approx&
f_{\psi (t-1)} (x_0) \cdot
\frac{(1-\omega_\psi)}{(\tilde{\omega}_{t-1} \mu_{t-1})} \cdot
\left( x_0 - \mu_{\psi (t-1)}\right)  \nonumber \\
&& \times \left( \frac{\tilde{\omega_t} \mu_t}{\tilde{\omega}_{t-1} \mu_{t-1}} -1 \right) \nonumber \\
&\approx&
f_{\psi (t-1)} (x_0) \cdot
\lambda_{\psi (t-1)} \cdot \left( x_0 - \mu_{\psi (t-1)}\right) \cdot
\left( \frac{\tilde{\omega_t} \mu_t}{\tilde{\omega}_{t-1} \mu_{t-1}} -1 \right) \nonumber \\
&&
\end{eqnarray}
(9) says that the forward difference, 
$f_{\psi t} (x_0) - f_{\psi (t-1)} (x_0)$, 
of relative frequencies of the same $x_0$ is proportional to 
$f_{\psi (t-1)} (x_0)$, the scale parameter at time $t-1$, 
$\lambda_{\psi (t-1)}$,
the signed difference between $x_0$ and the conditional mean, 
$\mu_{\psi (t-1)}$ 
at time $t-1$, and the signed proportional increase (positive) 
or proportional 
decrease (negative) in the product $(\tilde{\omega}_{t-1} \mu_{t-1})$. 
(9) implies little change in the relative frequency of wage
income in the vicinity of the conditional mean. It also implies that the 
$(x_0 - \mu_{\psi (t-1)})$ term can become largest in absolute value in the 
extreme right tail, i.e., for the largest $x_0$, 
since the absolute value of the 
difference $(x_0 - \mu_{\psi (t-1)})$ is greater for the
maximum $x_0$, typically more than three times the mean, than it is for the
minimum $x_0$, which is very nearly one mean away from the mean. However, the
forward difference, $(f_{\psi t} (x_0) - f_{\psi (t-1)} (x_0))$, 
will still be forced down toward zero in the far right tail when 
$(\tilde{\omega}_{t-1} \mu_{t-1})$ 
increases because the RHS of (9) is multiplied by $(f_{\psi (t-1)} (x_0))$ 
which becomes small quickly as $x_0$ becomes large. So the forward difference
becomes small in the far right tail even when 
$(\tilde{\omega}_{t-1} \mu_{t-1})$ 
increases despite the fact that $(x_0 - \mu_{\psi (t-1)})$ 
reaches its positive 
maximum for the maximum $x_0$. However, Newton's approximation to the ratio, 
$(f_{\psi t}(x_0)/f_{\psi (t-1)}(x_0))$, reflects the full effect of 
$(x_0 - \mu_{\psi (t-1)})$ on growth in the density of the 
far right tail when $(\tilde{\omega}_{t-1} \mu_{t-1})$ increases.

Given that in the IP's macro-model change is exogenous, due to 
$(\tilde{\omega_t} \mu_t)$, the ratio, $f_{\psi t}(x_0)/f_{\psi (t-1)}(x_0)$, 
is approximated via Newton's approximation as:
\begin{eqnarray}
\label{ja:fpsitoverfpsit-1}
\frac{f_{\psi t} (x_0)}{f_{\psi (t-1)} (x_0)}
&\approx&
\frac{f_{\psi (t-1)} (x_0) + f^\prime_{\psi (t-1)} (x_0) \left( (\tilde{\omega_t} \mu_t) - (\tilde{\omega}_{t-1} \mu_{t-1})\right)}{f_{\psi (t-1)} (x_0)} \nonumber \\
&\approx&
\left[ 1 + 
\left[ (x_0 - \mu_{\psi (t-1)})
\left( \frac{1-\omega_\psi}{\tilde{\omega}_{t-1} \mu_{t-1}} \right)
\left( \frac{\tilde{\omega_t} \mu_t}{\tilde{\omega}_{t-1} \mu_{t-1}} - 1\right)
\right] 
\right]
\end{eqnarray}
The bigger the $(x_0 - \mu_{\psi (t-1)})$ term is in the right tail, 
the greater is the ratio $f_{\psi t}(x_0)/f_{\psi (t-1)}(x_0)$ when 
$(\tilde{\omega_t} \mu_t)$ increases. Figure 1, showing the surge in 
wage \underline{income nouveaux} riches, graphs the empirical 
analogue of the ratio $f_{\psi t}(x_0)/f_{\psi (t-1)}(x_0)$. (10) is
descriptive of figure 1. Note that according to (10), in the right 
tail where $x_0 > \mu_{\psi (t-1)}$, the difference 
$(x_0 - \mu_{\psi (t-1)})$  for $x_0$ fixed becomes smaller as the 
conditional mean, $\mu_{\psi (t-1)}$, increases with increasing 
$(\tilde{\omega_t} \mu_t)$, implying a deceleration in the rate of
increase of the ratio $f_{\psi t}(x_0)/f_{\psi (t-1)}(x_0)$ 
for a given increase in $(\tilde{\omega_t} \mu_t)$. 
This deceleration is evident in figure 1.

The expression for forward proportional change is that of the RHS of 
(10) minus 1.0:
\begin{eqnarray}
\label{ja:fpsitfpsit-1}
\frac{f_{\psi t} (x_0) - f_{\psi (t-1)} (x_0)}{f_{\psi (t-1)} (x_0)}
&\approx&
\frac{f^\prime_{\psi (t-1)} (x_0) \left( (\tilde{\omega_t} \mu_t) - (\tilde{\omega}_{t-1} \mu_{t-1})\right)}{f_{\psi (t-1)} (x_0)} \nonumber \\
&\approx&
\left[ (x_0 - \mu_{\psi (t-1)})
\left( \frac{1-\omega_\psi}{\tilde{\omega}_{t-1} \mu_{t-1}} \right)
\left( \frac{\tilde{\omega_t} \mu_t}{\tilde{\omega}_{t-1} \mu_{t-1}} - 1\right)
\right] \nonumber \\
&\approx&
\lambda_{\psi (t-1)}  \left( x_0 - \mu_{\psi (t-1)}\right) 
\left( \frac{\tilde{\omega_t} \mu_t}{\tilde{\omega}_{t-1} \mu_{t-1}} -1 \right) 
\end{eqnarray}
and it has like (9) the property that it changes sign according 
to whether $x_0$ is greater than or less than the conditional mean,
$\mu_{\psi t}$, and whether $\tilde{\omega_t} \mu_t$ has increased 
or decreased. For example, in the right tail of the distribution, i.e., 
$x_0 > \mu_{\psi t}$, when $\tilde{\omega_t} \mu_t$
increases, forward proportional change in the distribution, 
$f_{\psi t}(x_0)$, is positive. Forward proportional change in 
the distribution is a product of the three factors on the RHS of (11). 
Forward proportional change in the distribution, $f_{\psi t}(x_0)$, 
in (11) is a linear function of the difference $(x_0 - \mu_{\psi (t-1)})$ 
and can, since maximum $x_0$ can be at least three times as far from 
the mean as minimum $x_0$, forward proportional growth in the extreme 
right of the right tail when $\tilde{\omega_t} \mu_t$ increases is 
greater than at any other income amount. In other words, the IP's 
macro model implies rapid growth in the population of wage income 
\underline{nouveaux riches} whenever $(\tilde{\omega_t} \mu_t)$ increases. 
One would expect that purveyors of goods and services priced 
for people with large wage incomes might see their market experiencing 
explosive growth whenever the product $(\tilde{\omega_t} \mu_t)$ increases.

\subsection{The Implied Dynamics of the IP's Macro Model for the 
Unconditional Distribution of Wage Income}
The IP's macro model of the unconditional wage income distribution,
a mixture of gamma pdf's, $f_t(x_0)$, is:
\begin{equation}
\label{ja:}
f_t(x_0) = u_{1t} f_{1t}(x_0) + \ldots + u_{\psi t} f_{\psi t}(x_0) + 
\ldots + u_{\Psi t} f_{\Psi t}(x_0)
\end{equation}
where:

\noindent
$f_{\psi t}(x_0) \equiv$  IP's macro model of distribution of wealth 
in the $\omega_\psi$ equivalence class at time $t$;

\noindent
$u_{\psi t} \equiv$ proportion of particles in the $\omega_\psi$ 
equivalence class at time $t$, the mixing weights.

The dynamics of (12), the unconditional relative frequency of wage income, 
are driven by $(\tilde{\omega_t} \mu_t)$ as in (7) and also by the
direct effect of the $u_{\psi t}$'s:
\begin{equation}
\label{ja:delft}
\frac{\partial f_t (x_0)}{\partial (\tilde{\omega_t} \mu_t)} =
\sum_\psi
\left(
u_{\psi t} \ f_{\psi t} (x_0) \ \frac{(1-\omega_\psi)}{(\tilde{\omega_t} \mu_t)^2} \ (x_0 -\mu_{\psi t})
\right)
\end{equation}

(12) is a gamma pdf mixture; a gamma mixture is not, in general, a
gamma pdf. While (12) shares many properties of (2) in the $\omega_\psi$ 
equivalence class, it has others as well, namely the direct effect of 
change in the proportions, the $u_{\psi t}$'s, in each $\omega_\psi$ 
equivalence class. Figure 12 shows that the $u_{\psi t}$'s of larger 
$\omega_\psi$'s (those of the less well educated, e.g. workers without 
a high school diploma) decreased between 1962 and 2004 while the 
$u_{\psi t}$'s of the smaller $\omega_\psi$'s (those of the
more educated, e.g., with at least some post-secondary school education)
increased. This change in the $u_{\psi t}$'s implies that 
$\tilde{\omega}_t$ decreased in this period, as figure 13 shows.

\begin{figure}
\centering
    {\resizebox{5.5cm}{!}{\includegraphics{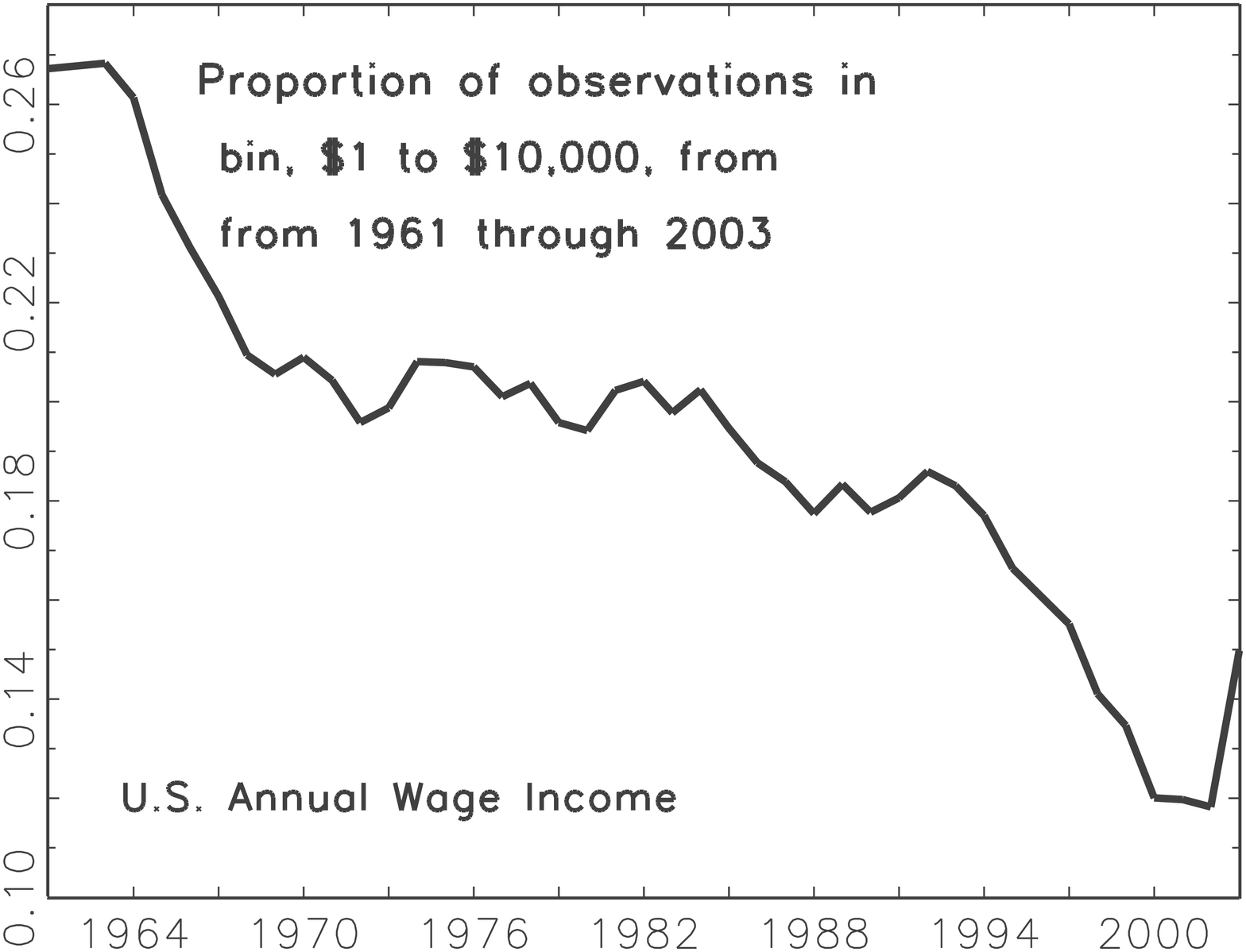}}}
    {\resizebox{5.5cm}{!}{\includegraphics{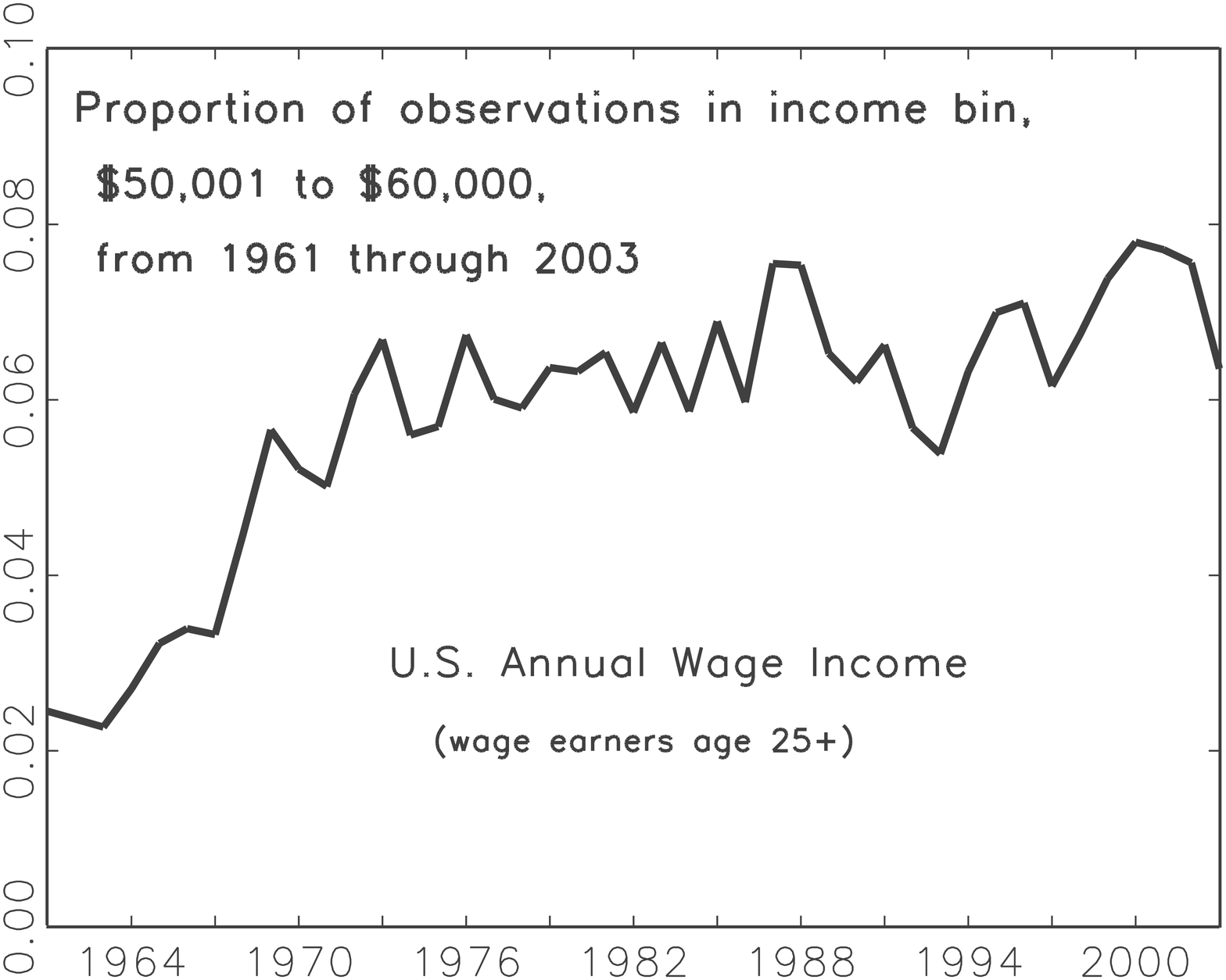}}}
\\    Fig. 17 \hskip 2in Fig. 18
\caption{\label{jafig:fig17}
Relative frequencies of incomes \$1 - \$10,000 in the unconditional
distribution.
Source: Author's estimates from data of the March Current Population Survey.
}
\caption{\label{jafig:fig18}
Relative frequencies of incomes
\$50,001 - \$60,000 in the unconditional distribution.
Source: Author's estimates from data of the March Current Population Survey.
}
\end{figure}
The implications for the right tail of the conditional distribution, 
$f_{\psi t} (x_0)$, in (9), (10), and (11), as the product 
$(\tilde{\omega_t} \mu_t)$ increases, carry through for the
dynamics of the right tail of the unconditional distribution, 
$f_{t} (x_0)$, for $x_0 > \mu_{\phi t}$ where $\mu_{\phi t}$ 
is the mean of $x$ in the $\omega_\phi$ equivalence class where 
$\omega_\phi$ is the minimum $\omega$, (and consequently $\mu_{\phi t}$ 
is the maximum mean of any $\omega$ equivalence class), and
for the dynamics of the left tail of the unconditional distribution, 
$f_t(x_0)$, for $x_0 < \mu_{\theta t}$ where $\mu_{\theta t}$ is the mean of 
$x$ in the $\omega_\theta$ equivalence class, where $\omega_\theta$ 
is the maximum $\omega$ in the population (and consequently $\mu_{\theta t}$ 
is the minimum mean of any $\omega$ equivalence class). Thus, as 
$(\tilde{\omega_t} \mu_t)$ and the $u_\psi$ in equivalence classes with
smaller $\omega_\psi$'s increase, (13) implies that the left tail 
thins and the right tail thickens. Figures 17 and 18 show that such 
is the case in the left tail bin, \$1-\$10,000, and the right tail bin, 
\$50,001 - \$60,000 (both in constant 2003 dollars).
Figure 19 shows how each relative frequency (that in the bin 
\$1-\$10,000 and that in the bin, \$50,001-\$60,000) has a large 
positive correlation with other relative frequencies in the same 
tail and a large negative correlation with relative
frequencies in the other tail. For example, the relative 
frequencies in the bins \$1 - \$10,000 and \$50,001 - \$60,000 
have a nearly a perfect negative correlation with
each other. Both relative frequencies, as one would expect 
given (13), have a near zero correlation with relative 
frequencies close to the unconditional mean of wage income.
\begin{figure}
\centering
    {\resizebox{6.0cm}{!}{\includegraphics{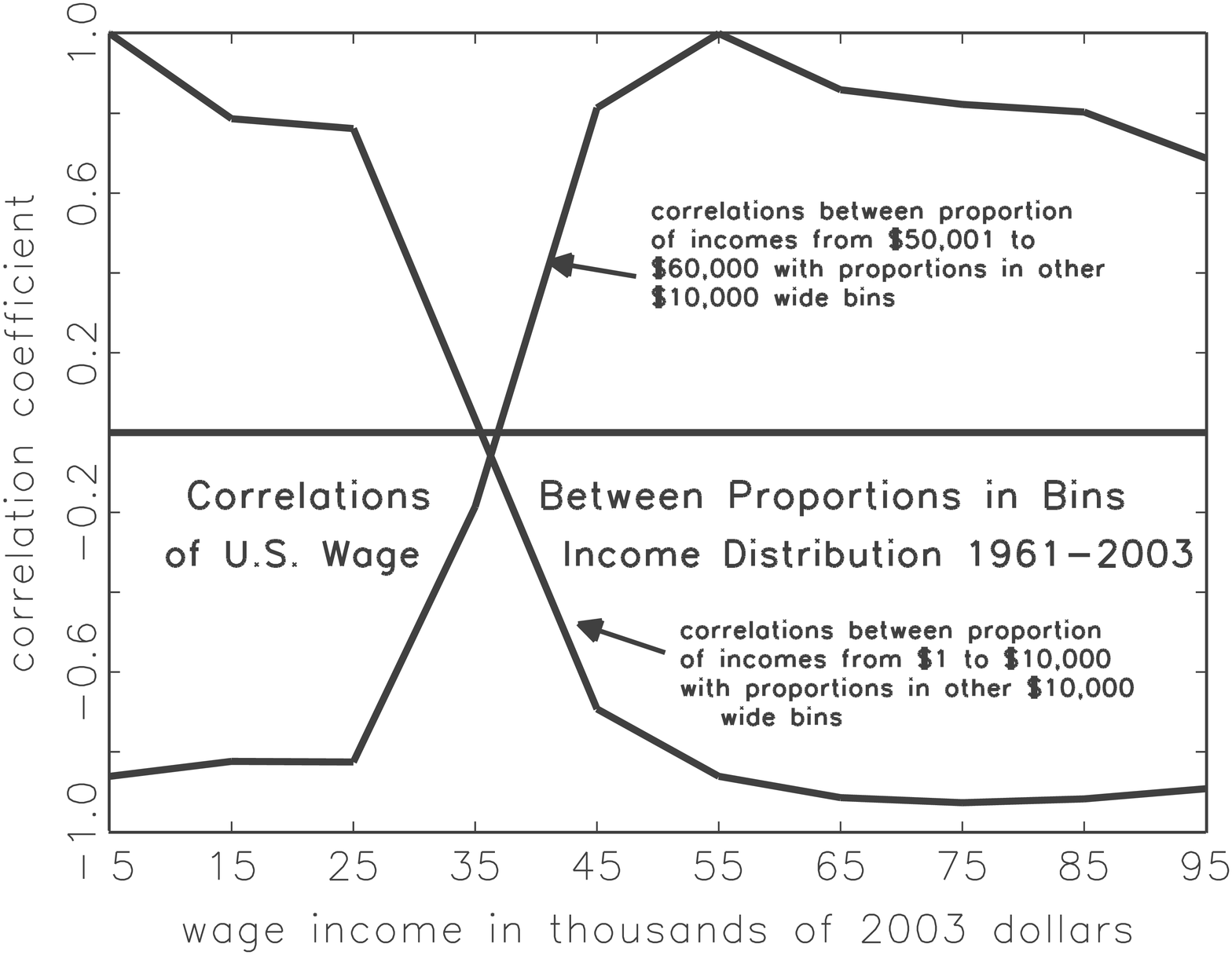}}}
    {\resizebox{6.0cm}{!}{\includegraphics{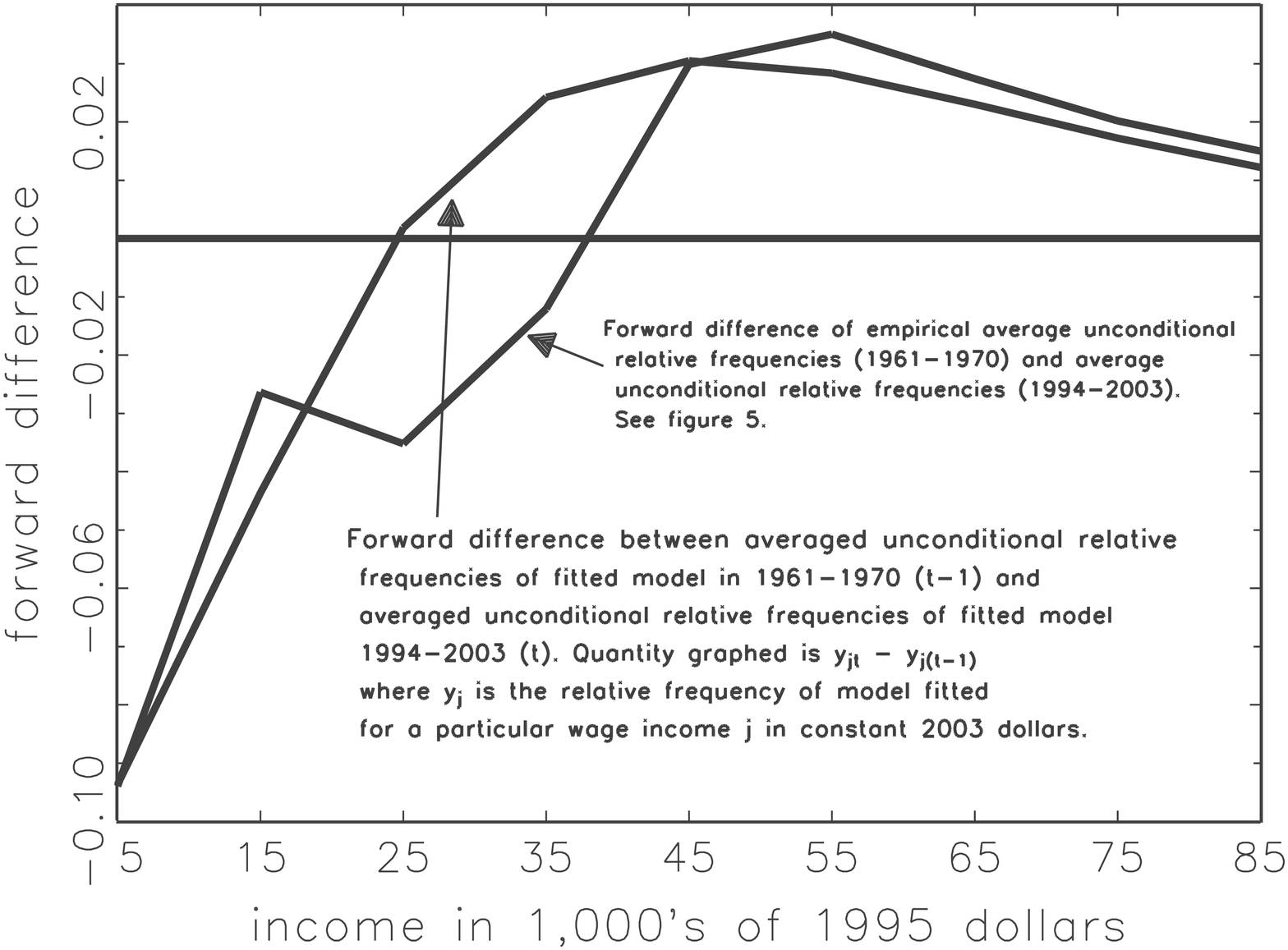}}}
\\    Fig. 19 \hskip 2in Fig. 20
\caption{\label{jafig:fig19}
Correlations between relative
frequency in wage income bin in left tail, \$1,-\$10,000, 
and relative frequency in right tail bin,
\$50,001 -\$60,000, and relative frequencies in all
the other income bins around the distribution.
Source: Author's estimates from data of the March Current Population Survey.
}
\caption{\label{jafig:fig20}
Source: Author's estimates from data of the March Current Population Survey.
}
\end{figure}

Figure 20 shows that the unconditional forward difference of
wage incomes between the average of the relative frequencies in the
period 1961 to 1970 and the average of the relative frequencies in the
period 1994-2003 largely overlaps the fitted forward difference
between the expected relative frequencies in these two periods at
the beginning and end of the time series. The time averaging is done
to smooth out the pronounced frequency spiking in these data. See
Angle (1994) for a discussion of
frequency spiking in the wage income observations collected by the March
Current Population Survey.

\section{Conclusions}
The IP's macro model fits the distribution of U.S. wage income 
conditioned on education 1961-2003. It also accounts
for one of the quirkier time-series of scalar statistics of U.S. 
wage income in the same time period: the more rapid growth in
the relative frequency of the larger wage income in the right tail 
of the distribution, that is, among wage incomes greater than
mean wage income. Figure 20 shows that the IP's macro model accounts 
for how the relative frequencies of wage income
changed between 1961-1970 and 1994-2003. Figure 21 shows why, 
in particular, for large wage incomes (defined in constant
2003 dollars): the expected frequencies of large wage incomes 
under the IP's macro model track the observed frequencies of
large wage incomes closely.

The observed relative frequencies are estimated from reports of 
personal annual wage income in the micro-data file,
the individual person records, of the March Current Population 
Survey (CPS), in `public use' form, i.e., with personal
identifiers stripped from the file. The March CPS is a survey 
of a large sample of households in the U.S. conducted by the U.S.
Bureau of the Census. In the March CPS, a respondent answers questions 
posed by a Census Bureau interviewer about members
of the household. There is a question about the annual wage income 
of each member of the household in the previous year. See
figure 4 for estimates of the distribution of annual wage income 
1961-2003. All dollar amounts have been converted to constant
2003 dollars.

The U.S. Census Bureau has evaluated the adequacy of its wage 
income question in the March Current Population Survey
(CPS) and acknowledged that respondents, on average, underestimate 
the wage income they report (Roemer, 2000: 1). Roemer
writes ``Many people are reluctant to reveal their incomes to survey 
researchers and this reluctance makes such surveys
particularly prone to response errors.'' Roemer (2000: 17-21) 
reports that these underestimates are least biased downward for
wage incomes near the median but seriously biased downward for 
large wage incomes. So it is not a problem for the IP's macro
model if it overestimates the relative frequency of large wage 
incomes slightly, particularly very large wage incomes, as you can
see it does in figure 21.
\begin{figure}
\centering
    {\resizebox{6.0cm}{!}{%
       \includegraphics{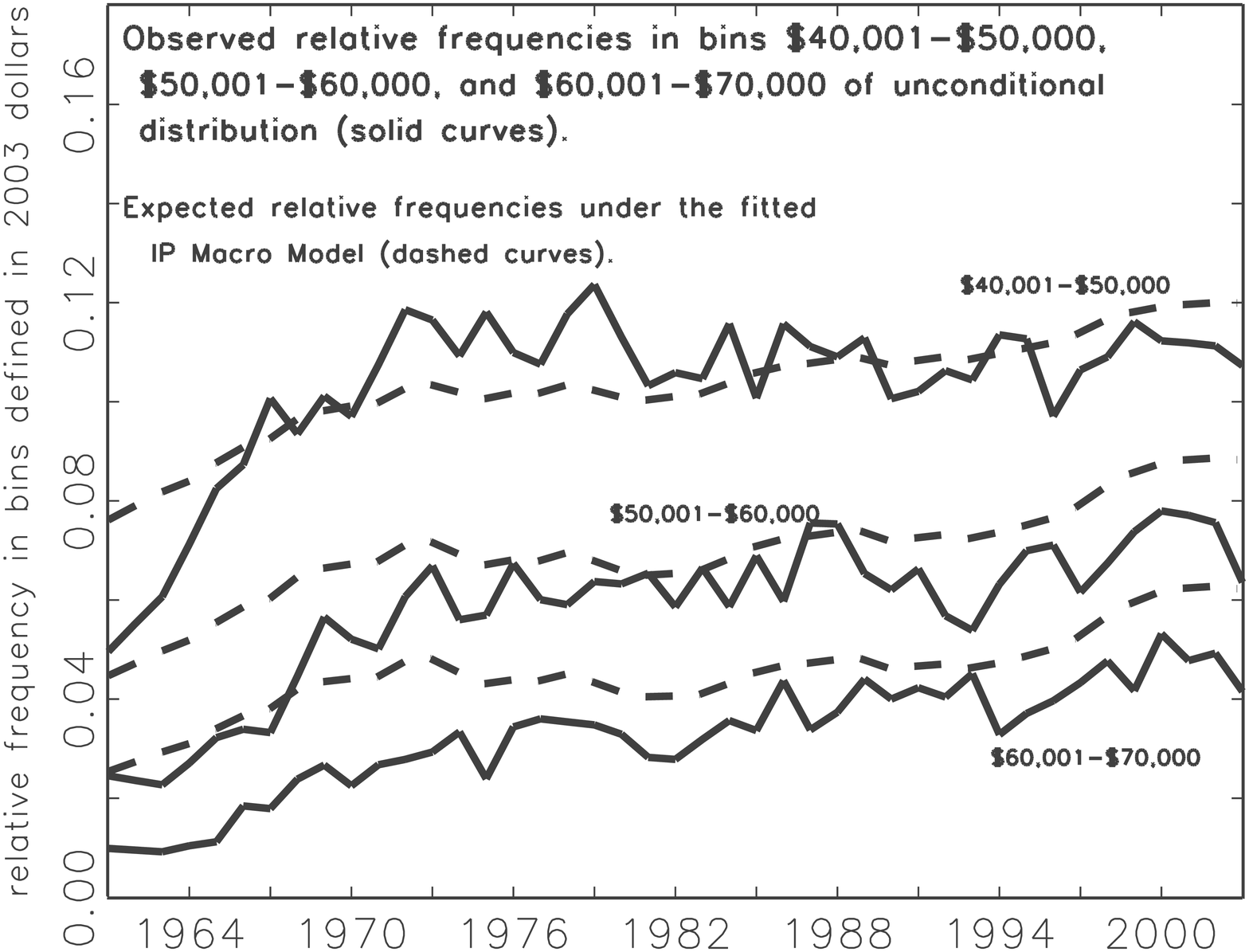}
}}
\caption{\label{jafig:fig21}
Unconditional right tail relative frequencies 1961-2003 (solid curves) and
estimated right tail relative frequencies under
the fitted IP Macro Model (dashed curves).
Source: Author's estimates from data of the March Current Population Survey.
}
\end{figure}

The macro model of the Inequality Process is a gamma probability 
density function (pdf) whose parameters are derived
from the micro model of the Inequality Process and expressed in 
terms of its parameters. See (2) through (5). The dynamics of
this model are expressed in terms of the gamma scale parameter, 
(5), of this model. (5) says that the model is driven
exogenously by the product $(\tilde{\omega_t} \mu_t)$ through its 
scale parameter, $\lambda_{\psi t}$. $(\tilde{\omega_t} \mu_t)$ 
is a function of the distribution of education in the
labor force at time $t$ and the unconditional mean of wage income, 
$\mu_t$, at time $t$. $\tilde{\omega_t}$ is the harmonic mean of the 
estimated IP parameters, the $\omega_\psi$'s. These are estimated 
in the fitting of the IP's macro model to the distribution of wage 
income conditioned on education, 1961-2003. The $\omega_\psi$'s 
also enter the 
formula by which the unconditional mean, $\mu_t$, is estimated from sample
conditional medians under the hypothesis that wage income is gamma 
distributed. The $u_\psi$'s, the proportions in each $\omega_\psi$
equivalence class, by hypothesis the fraction of the labor force at 
a particular level of education, also enter the formula by which
$\mu_t$ is estimated from sample conditional medians. The IP's 
macro model fits the distribution of wage income in the U.S., 1961-2003, 
well.

\subsection{The Dynamics of the Wage Income Distribution When 
$(\tilde{\omega_t} \mu_t)$ Increases: A Stretching, Not a `Hollowing Out'}
\subsubsection{Not a `Hollowing Out'}
When $(\tilde{\omega_t} \mu_t)$ increases, the distribution of wage 
income stretches to the right over larger wage incomes, as in the
comparison of figure 8 to figure 7 . Figure 8 is the graph of gamma 
pdf's with different shapes but the same scale parameter.
Figure 8 has gamma pdfs with the same shape parameters but a different 
scale parameter, one that is half that of figure 7. The
gamma pdf's of figure 8 look stretched to the right. When 
$(\tilde{\omega_t} \mu_t)$ decreases, the wage income distribution is 
compressed to the left over smaller wage incomes, as in the comparison 
of figure 9 to figure 7. These effects are deduced from the IP's macro model 
in (9). The last term in the product on the RHS of (9) is positive when
$(\tilde{\omega_t} \mu_t) > (\tilde{\omega}_{t-1} \mu_{t-1})$, negative when 
$(\tilde{\omega_t} \mu_t) < (\tilde{\omega}_{t-1} \mu_{t-1})$, meaning that when 
$(\tilde{\omega_t} \mu_t)$ increases, the right tail thickens, the left tail
thins, and vice versa when $(\tilde{\omega_t} \mu_t)$ decreases. While the IP's 
micro model is time-reversal asymmetric, its macro model is
time-reversal symmetric. The IP's macro model implies in (10), and (11) 
that growth in the relative frequency of large wage
incomes, i.e., the thickness of the right tail of the wage income 
distribution, is greater, the larger the wage income, i.e., the
farther to the right in the tail, when $(\tilde{\omega_t} \mu_t)$ increases.

So the IP's macro model accounts for the surge in the far right tail 
of the wage income distribution in the U.S., the appearance of wage 
income \underline{nouveaux riches}, as $(\tilde{\omega_t} \mu_t)$ 
increased from 1961 through 2003. See figures 1, 5, 20, and 21. The
IP's macro model implies that the right tail of the wage income 
distribution thickened as $(\tilde{\omega_t} \mu_t)$ increased from 
1961 through 2003 and the left tail of the distribution thinned. 
The empirical evidence bears out this implication of the IP's macro 
model, but contradicts the interpretation in the labor economics 
literature that the thickening of the right tail of the wage income
distribution represented a `hollowing out' of the wage income 
distribution, that is, a simultaneous thickening in the left and
right tails of the distribution at the expense of the relative frequency 
of wage incomes near the median of the distribution, as
illustrated conceptually in figure 3.

As you can see in figure 4, the unconditional distribution of wage income 
thinned in its left tail and thickened in its right
from 1961 through 2003. Figure 17 shows how the relative frequency of 
wage incomes from \$1- \$10,000 (constant 2003
dollars) decreased from 1961 through 2003, although not monotonically, 
while figure 18 shows how the relative frequency of
wage incomes from \$50,001 - \$60,000 (constant 2003 dollars) increased 
from 1961 through 2003, although not monotonically.
$\$50,001$ in 2003 dollars is greater than the unconditional mean of 
wage income from 1961 through 2003, so the wage income
bin \$50,001-\$60,000 was in the right tail the entire time. If there 
is any remaining question of what was happening elsewhere
in the distribution, it is answered by figure 19 which shows the 
correlation between the relative frequency in income bin
\$1-\$10,000 with that of every other income bin. The relative 
frequency of this extreme left tail bin was positively correlated
with the relative frequency in the other left tail bin, had almost no 
correlation with relative frequency of mean wage income,
and a large negative correlation with relative frequencies of all 
the right tail income bins. Figure 19 also shows the correlations
of the relative frequency of the income bin \$50,001 - \$60,000 with 
relative frequencies in other bins around the distribution.
These correlations are a near mirror image of the correlations of the 
left tail bin \$1 - \$10,000. The relative frequency of income
bin \$50,001 - \$60,000 has a high positive correlation with the 
relative frequencies of other right tail income bins, near zero
correlation with the relative frequency of mean income, and a large 
negative correlation with the relative frequencies of left tail
wage income bins. Figure 20 shows that the relative frequency of wage 
incomes smaller than the mean decreased between 1961
and 2003 while those greater than the mean increased. There is no 
doubt that the relative frequencies of the left tail of the wage
income distribution vary inversely with the relative frequencies of 
the right tail, just as the IP's macro model implies, in
contradiction of the `hollowing out' hypothesis.

\subsubsection{A Stretching of the Distribution When 
$(\tilde{\omega_t} \mu_t)$ Increases}

This paper has focused on how the relative frequency of a wage income 
of a given size changes when $(\tilde{\omega_t} \mu_t)$ increases
because it is algebraically transparent. The algebra indicates more 
rapid growth in the relative frequency of the larger wage
income in the right tail of the distribution. However, a clearer 
demonstration of how the IP's macro model and the empirical
wage income change when $(\tilde{\omega_t} \mu_t)$ increases is in the 
dynamics of the percentiles of wage income, that is, not how the relative
frequency of a particular fixed wage income in constant dollars, $x_0$, 
changes, but rather how the percentiles of the distribution
change. Figure 2 shows that the 90th percentile of wage income increased 
more in absolute terms than the 10th percentile
between 1961 and 2003, i.e., the distribution stretched farther to the 
right over larger wage incomes in its right tail than its left.
Does the same occur with the 10th and 90th percentiles of the IP's macro 
model of the unconditional distribution of wage
income? This demonstration requires numerical integration and so 
is less transparent algebraically than inspecting the algebra
of the model for the dynamics of the relative frequency of large wage incomes.

Figure 22 displays how well the percentiles of the model track the 
observed percentiles of wage income. The tendency
to slightly overestimate the 90th percentile is not a problem given 
Roemer's (2000) evaluation of the accuracy of reporting of
wage income data in the March CPS. In figure 22 the graphs of the 
unconditional percentiles of the IP's macro model and of
empirical wage income as $(\tilde{\omega_t} \mu_t)$ increases show 
both distributions stretching to the right: the bigger the percentile, 
the more it increases in absolute constant dollars, what one would 
expect from the multiplication of all wage income percentiles by the
same multiplicative constant, usually greater than 1.0, in each year 
between 1961 and 2006.

A percentile, $x_{(i)\psi t}$, of the IP's macro model, 
$f_{\psi t} (x)$, is:
\[ \frac{i}{100} =
\int_0^{x_{(i) \psi t}} \frac{\lambda_{\psi t}^{\alpha_\psi}}{\Gamma (\alpha_\psi)}
x^{\alpha_\psi -1} \exp(- \lambda_{\psi t} x) \ dx 
\]
where $i$ is integer and $i$ is less than or equal to 100. 
Figure 15 graphs the conditional medians, the 50th percentiles, 
$x_{(50)\psi t}$'s, from
1961 through 2003. Figure 16 shows that, when standardized, i.e., 
when their mean is subtracted from them and this difference
is divided by their standard deviation, the transformed conditional 
medians have a time-series close to that of the
standardization of $(\tilde{\omega_t} \mu_t)$. (6), Doodson's 
approximation to the median of a gamma pdf in terms of the IP's 
parameters, shows why: the median is approximately a constant 
function of $(\tilde{\omega_t} \mu_t)$. $(\tilde{\omega_t} \mu_t)$ 
enters $f_{\psi t}(x)$ as a gamma scale parameter
transformation, via $\lambda_{\psi t}$. A scale transformation affects 
all percentiles multiplicatively, as in the comparison of figure 8 to figure
7. The gamma pdfs of figure 8 have the same shape parameters as 
those of figure 7. The difference between the two sets of
graphs is that those of figure 8 have scale parameters, $\lambda_{\psi t}$, 
that are one half those of figure 7. The gamma pdfs of figure 8 have
been stretched to the right over larger $x$'s from where they were in 
figure 7. The IP's macro model implies this stretching to the
right over larger wage incomes when the product $(\tilde{\omega_t} \mu_t)$ 
increases, which figure 14 shows it did from 1961 through 2003,
although not monotonically so. A larger $(\tilde{\omega_t} \mu_t)$ 
results in a smaller gamma scale parameter, $\lambda_{\psi t}$, given (5).

So, the Inequality Process' (IP) macro model explains both the 
surge in the relative frequency of large wage incomes
and the greater absolute increase in the greater percentile of wage incomes 
in the U.S., 1961-2003 as $(\tilde{\omega_t} \mu_t)$ increased. Since
the $\tilde{\omega_t}$ term decreases with rising levels of 
education in the U.S. 
labor force, the condition of $(\tilde{\omega_t} \mu_t)$ increasing means that
the unconditional mean of wage income, $\mu_t$, grew more proportionally 
1961-2003 than $\tilde{\omega_t}$ decreased. Since all percentiles of
wage income grew as $(\tilde{\omega_t} \mu_t)$ increased, the surge in wage 
income \underline{nouveaux riches} in the U.S. 1961-2003 was simply a visible
indicator of generally rising wage incomes, hardly the ominous event it 
was made out to be by some in the scholarly literature and the popular press.

\subsubsection{Appendix A: The March Current 
Population Survey And Its Analysis}

The distribution of annual wage and salary income is estimated 
with data from the March Current Population Surveys (CPS)
(1962-2002), conducted by the U.S. Bureau of the Census. 
One of the money income questions asked on the March CPS
is total wage and salary income received in the previous calendar year. 
See Weinberg, Nelson, Roemer, and Welniak (1999) for
a description of the CPS and its history. The CPS has a substantial 
number of households in its nationwide sample. The March
Current Population Survey (CPS) provides the data for official 
U.S. government estimates of inequality of wage income as well
as most of the labor economics literature on inequality of wage income in the U.S.

The present paper examines the civilian population of the U.S. that is 25+ 
in age and earns at least \$1 (nominal) in annual
wage income. The age restriction to 25+ is to allow the more 
educated to be compared to the less educated. It is a
conventional restriction in studies of the relationship of education 
to wage income. The data of the March CPS of 1962 through
2004 were purchased from Unicon Research, inc. (Unicon Research, inc, 2004; 
Current Population Surveys, March
1962-2004), which provides the services of data cleaning, documentation 
of variable definitions and variable comparability
over time, and data extraction software. Unicon Research, inc was 
not able to find a copy of the March 1963 CPS public use
sample containing data on education. Consequently, the distribution 
of wage and salary income received in 1962 (from the
March 1963 CPS) conditioned on education is interpolated from the 
1961 and 1963 (from the 1962 and 1964 March CPS').

All dollar amounts in the March CPS are converted to constant 2003 
dollars using the U.S. Bureau of Economic Analysis
National Income and Product Account Table 2.4.4 Price indexes 
for personal consumption expenditure by type of product
[index numbers, 2000 = 100] 

\noindent
http://www.bea.gov/bea/dn/nipaweb/TableView.asp\#Mid 
[Last revised on 8/4/05].

\subsubsection{Appendix B: Estimation}

\noindent
\textbf{\textit{Estimation of Relative Frequencies}}

\noindent
All estimates are weighted estimates. The weight associated with the 
$j$th observation in the $t$th year, $u_{jt}^*$, is:
\[
u_{jt}^* = \frac{u_{jt}}{\sum_{i=1}^{n_t} u_{it}} \ n_t
\]
where,

$u_{jt}$ = the raw weight provided by the Census Bureau for observation $j$

$n_t$ = the sample size in year $t$.

\vskip .1in
\noindent
\textbf{\textit{Estimation of the $\mu_t$, the Unconditional Mean, 
from Sample Conditional Medians, $x_{(50)\psi t}$'s}}

\noindent
While an unconditional sample mean of wage incomes in the March CPS can 
be directly estimated from the data, it is
known to be an underestimate of the population unconditional mean, $\mu_t$. 
The sampling frame of the March CPS does not
sample large wage incomes at a higher rate than smaller wage incomes. 
Consequently, given the right skew of the distribution
wage income dollars will be missed in the form of very large individual 
wage incomes biasing the sample mean of wage income
downward. Further, the Census Bureau itself has concluded that even 
when a household with one or more large wage incomes
falls into the sample, those wage income reports have a serious 
downward bias (Roemer, 2000:17-21). The sample median of
wage incomes is robust against these problems of estimation. It is 
as well measured as any sample statistic of annual wage income.

The unconditional mean of the IP's macro model, $\mu_t$, 
is estimated in terms of the sample conditional medians, the
$x_{(50)\psi t}$'s, (the median wage income at the $\psi$th level of 
education) and the $u_{\psi t}$'s, (the proportion of the labor 
force at the $\psi$th level of education) using Doodson's approximation 
formula for the median of a gamma pdf, (Weatherburn, 1947:15 [cited in Salem
and Mount, 1974]) as instantiated for the IP's macro model in (6), since:
\[
\mu_t = u_{1t} \mu_{1t} + u_{2t} \mu_{2t} + \ldots + u_{\psi t} \mu_{\psi t}
+ \ldots + u_{6t} \mu_{6t}.
\]

\end{document}